\documentclass[10pt,twocolumn,letterpaper]{article}
\usepackage[pagenumbers]{cvpr} % To force page numbers, e.g. for an arXiv version

\usepackage{algorithmic}
\usepackage[linesnumbered,ruled]{algorithm2e}
\usepackage{color}
\usepackage{graphicx}
\usepackage{amsmath}
\usepackage{amssymb}
\usepackage{booktabs}
\usepackage{dsfont}
\usepackage{multirow}
\usepackage{subcaption}
\usepackage[misc]{ifsym}
\usepackage{enumitem}
\usepackage[pagebackref,breaklinks,colorlinks]{hyperref}

\usepackage[capitalize]{cleveref}
\crefname{section}{Sec.}{Secs.}
\Crefname{section}{Section}{Sections}
\Crefname{table}{Table}{Tables}
\crefname{table}{Tab.}{Tabs.}

\def\confName{CVPR}
\def\confYear{2022}

\newtheorem{definition}{Definition}
\DeclareMathOperator*{\argmax}{arg\,max}
\begin{document}

%%%%%%%%%%%%%%%%%%%%%%%%%%%%%%%%%%%%%%%%%%%%%%%%%%%%%%%%%%%%%%%%%%%%%
%  Commands for comments
\definecolor{dkgreen}{rgb}{0,0.6,0}
\definecolor{dkself}{rgb}{0.6,0.2,0.6}
\definecolor{dkpink}{rgb}{0.8,0.4,0.4}
\definecolor{dkred}{rgb}{0.6,0.0,0.0}

\newcommand{\shao}[1]{\textcolor{dkself}{[shaofeng: #1]}}
\newcommand{\gx}[1]{\textcolor{green}{\{\textbf{guoxing:} {\em#1}\}}}
\newcommand\mypara[1]{\vspace{1.0mm}\noindent\textbf{#1}}

% Quick Fix to capitalize Section reference in AutoRef
\def\sectionautorefname{Section}
\def\subsectionautorefname{Section}
\def\equationautorefname{Eq.}
\def\algorithmcautorefname{Alg.}
%%%%%%%%%%%%%%%%%%%%%%%%%%%%%%%%%%%%%%%%%%%%%%%%%%%%%%%%%%%%%%%%%%%%%
\newcommand{\norm}[1]{\left\lVert#1\right\rVert}
\newcommand{\secref}[1]{\mbox{Sec.~\ref{#1}}\xspace}
\newcommand{\secrefs}[2]{\mbox{Sec.~\ref{#1}--\ref{#2}}\xspace}
\newcommand{\figref}[1]{\mbox{Fig.~\ref{#1}}}
\newcommand{\tabref}[1]{\mbox{Table~\ref{#1}}}
\newcommand{\lstref}[1]{\mbox{Listing~\ref{#1}}}
\newcommand{\appref}[1]{\mbox{Appendix~\ref{#1}}}
\newcommand{\modelsymbol}[1]{f#1}
\newcommand{\vmodel}[1]{\modelsymbol{}_{\mathcal{V},#1}}
\newcommand{\pmodel}[1]{\modelsymbol{}_{\mathcal{P},#1}}
\newcommand{\amodel}[1]{\modelsymbol{}_{\mathcal{A},#1}}
\newcommand{\smodel}[1]{\modelsymbol{}_{\mathcal{S}#1}}
\newcommand{\fpfunc}{\mathcal{F}}
\newcommand{\fplen}{n}
\newcommand{\vdata}[1]{D_{\mathcal{V},#1}}
\newcommand{\pdata}[1]{D_{\mathcal{P},#1}}
\newcommand{\cldata}[1]{D_{\mathcal{C},#1}}
\newcommand{\xdomain}{\mathbb{R}^M}
\newcommand{\ydomain}{\mathbb{R}^N}

\newcommand{\uap}[1]{\mathbf{v}#1}

\title{Fingerprinting Deep Neural Networks Globally via Universal\\ Adversarial Perturbations

}
\author{Zirui Peng\textsuperscript{1}\thanks{Equal contribution.}\quad Shaofeng Li\textsuperscript{1}\footnotemark[1]\quad Guoxing Chen\textsuperscript{1, }$^{\textrm{\Letter}}$\quad Cheng Zhang\textsuperscript{2}\quad Haojin Zhu\textsuperscript{1, }$^{\textrm{\Letter}}$\quad Minhui Xue\textsuperscript{3,}\textsuperscript{4} \vspace{0.5em} \\
\normalsize{
\textsuperscript{1 }Shanghai Jiao Tong University \hspace{1em} \textsuperscript{2 }The Ohio State University \hspace{1em}
\textsuperscript{3 }CSIRO's Data61 \hspace{1em}
\textsuperscript{4 }The University of Adelaide \vspace{0em} 
}
\\
{\footnotesize \texttt{\{pengzirui,shaofengli,guoxingchen,zhu-hj\}@sjtu.edu.cn, zhang.7804@osu.edu, jason.xue@adelaide.edu.au}} 
}

\maketitle

%%%%%%%%% ABSTRACT
\begin{abstract}
In this paper, we propose a novel and practical mechanism to enable the service provider to verify whether a suspect model is stolen from the victim model via model extraction attacks. 
Our key insight is that the profile of a DNN model's decision boundary can be uniquely characterized by its \emph{Universal Adversarial Perturbations (UAPs)}. UAPs belong to a low-dimensional subspace and piracy models' subspaces are more consistent with victim model's subspace compared with non-piracy model.
Based on this, we propose a UAP fingerprinting method for DNN models and train an encoder via \textit{contrastive learning} that takes fingerprints as inputs, outputs a similarity score.
Extensive studies show that our framework can detect model Intellectual Property (IP) breaches with confidence $>$ 99.99 \% within only 20 fingerprints of the suspect model. It also has good generalizability across different model architectures and is robust against post-modifications on stolen models. 

\end{abstract}

%%%%%%%%% BODY TEXT
\section{Introduction}
In the past few years, deep learning has emerged as a promising approach and the foundation for a wide range of real-world applications. 
As network architecture becomes more and more sophisticated and training costs rise, well-trained models become lucrative targets for the adversary looking to ``steal'' them.
By querying the publicly available APIs of these models, an adversary can collect the outputs to train a piracy model, dubbed \textit{model extraction} attacks~\cite{DBLP:conf/uss/TramerZJRR16, DBLP:conf/uss/JagielskiCBKP20, DBLP:conf/uss/ChandrasekaranC20, DBLP:conf/crypto/CarliniJM20, DBLP:conf/icml/RolnickK20, DBLP:conf/icml/BeguelinTPK21, DBLP:conf/uss/ZhuCZL21}.

Existing works on mitigating model extraction attacks and protecting the Intellectual Property (IP) of trained models fall into two groups~\cite{DBLP:conf/sp/JiaYCDTCP21, DBLP:journals/corr/abs-2109-10870}. 
The first group is based on \textit{watermarking} techniques~\cite{DBLP:conf/uss/AdiBCPK18, DBLP:conf/uss/JiaCCP21, DBLP:conf/mm/SzyllerAMA21, DBLP:conf/cvpr/OngCNFY21, DBLP:journals/corr/abs-1809-00615, DBLP:journals/nca/MerrerPT20, DBLP:conf/ih/ShafieinejadLWL21}.
The idea is that the model owner introduces into her IP model a backdoor (\ie, a watermark), which would persist during the model extraction. By checking whether a suspect model contains the injected watermark, a defender can determine whether this model is pirated. 

The other category is based on \textit{fingerprinting} techniques that leverage inherent information (\ie, \textit{decision boundary}) of the trained models. Observing that a DNN model can be uniquely profiled by its decision boundary which is also likely to be inherited by piracy models, a model extraction attack can be identified by examining whether a suspect model has (almost) the same decision boundary as the victim model. A line of research~\cite{DBLP:conf/cvpr/HeZL19, DBLP:conf/iclr/LukasZK21, DBLP:conf/asiaccs/CaoJG21} adopts adversarial examples to represent the decision boundaries.

\begin{figure*}[thp]
    \centering
    \includegraphics[width=0.9\linewidth]{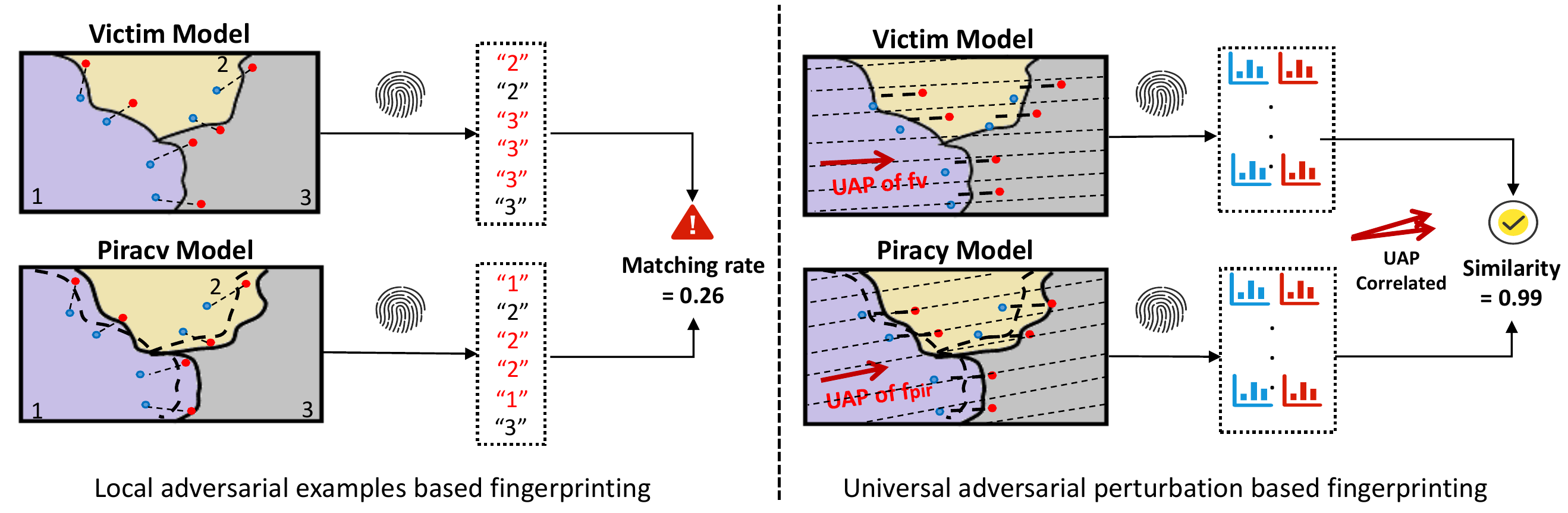}
    \vspace{-3mm}
    \caption{\small \textbf{Illustrations of local and universal adversarial perturbations.} Left: local adversarial perturbations are less robust to point-to-point decision boundary modification due to extraction. Right: our framework relies on the stable correlation of decision boundaries profiled by universal adversarial perturbations (UAPs).}
    \vspace{-3mm}
    \label{fig:my_label}
\end{figure*}
However, the effectiveness of existing mitigation schemes were challenged. Watermarking based solutions suffer from utility drop caused by watermarks. Another concern is that an attacker can illegally inject a backdoor to argue the ownership which violates the non-forgeable demand~\cite{DBLP:conf/uss/JiaCCP21}. Adversarial examples can only capture the \textit{local geometry}, particularly, orientations of the decision boundary in local regions surrounding the adversarial examples, which may fail to be transferred to the suspect model due to decision boundary variation during the extraction~\cite{DBLP:conf/ccs/PapernotMGJCS17,DBLP:conf/cvpr/KhrulkovO18}.

In this paper, we explore methods to capture the global characteristics of the decision boundary. As demonstrated in \autoref{fig:my_label}, we propose a more effective model extraction detection scheme based on \textit{Universal Adversarial Perturbations} (UAPs)~\cite{DBLP:conf/cvpr/Moosavi-Dezfooli17}. A carefully selected UAP vector $\uap$ can fool the  model on almost all datapoints. We find that UAPs are drawn from a low-dimensional subspace that contains most of the normal vectors of decision boundary. Due to decision boundary dependency, UAP subspaces of piracy models are more consistent with that of the victim model, which enables us to give a similarity score.   

There are two challenges in applying UAPs for detecting model extraction.
First, since the calculation of UAPs usually requires the knowledge of model parameters (\ie, white-box access) which model owners are not willing to provide, it is intractable for the defender to obtain UAPs of suspect models via black-box access. The second challenge is how to reliably distinguish between a piracy model and a homologous model (\ie, model trained on the same training data rather than the victim model's outputs and should be not considered ``stolen'').

To tackle the first challenge, we propose a fingerprinting function which is obtained by querying the suspect model with a number of datapoints, added by victim's UAPs. A more informative fingerprints need to capture as many parts of decision boundaries as possible. We therefore adopt K-means clustering on the last layer of the victim model to ensure that the datapoints are uniformly selected from different source classes and move towards different target classes. 

To address the second challenge, we design an encoder to map fingerprints of the victim model, piracy models, and homologous models into a joint representation space. 
We adopt \textit{contrastive learning}~\cite{DBLP:conf/icml/ChenK0H20} (which aims to shorten the distances of the samples in the same classes and push away any samples of other classes) to project homologous models farther away from the victim model than the piracy models.

In summary, we propose a more accurate, robust and general IP protection framework against the model extraction attacks. Our main contributions are:
\begin{itemize}[itemsep=0pt,topsep=2pt,leftmargin=12pt]
    \item We present one of the first attempts to leverage UAP distribution dependency to measure the decision boundary similarity between models. 
    We show that UAP outperforms adversarial perturbation for model fingerprinting.
    
    \item We propose a novel model ownership verification framework based on UAP fingerprinting that achieves a highly competitive detection rate in terms of AUC.

    \item Compared with prior fingerprinting works, we demonstrate the capability of our framework for detecting post-modificated piracy models.
    
    \item We adopt contrastive learning in encoder training to address the similarity gap between homologous models and piracy models. A new data augmentation approach is proposed to create ``views" for fingerprints.
\end{itemize}

\section{Background and Related Work}
\mypara{Model extraction}
violates the confidentiality of machine learning models~\cite{DBLP:conf/uss/TramerZJRR16, DBLP:conf/uss/JagielskiCBKP20, DBLP:conf/uss/ChandrasekaranC20, DBLP:conf/crypto/CarliniJM20, DBLP:conf/eurosp/JuutiSMA19}. 
In a model extraction attack, the attacker only has black-box access to a victim model and aims at stealing it through posing queries. 
The obtained model is expected to be functionally similar.
To extract a model, the attacker first needs to collect a set of unlabeled natural data. Then the natural data are mixed with carefully crafted synthesized data to query the victim model. 
The returned labels are then used to train a piracy model. This process repeats for several iterations until the piracy model recovers a satisfying utility of the victim.

\mypara{Model fingerprinting}
relies on finding existing features that characterises the model. Recent works rely on the different transferabilities of the victim model's adversarial examples~\cite{DBLP:conf/iclr/LiuCLS17, DBLP:journals/corr/TramerPGBM17, libo_2, bai2021ai, cezhang_2, DBLP:conf/iclr/MadryMSTV18} on piracy models and independent models. 
They are widely used to solve problems like model modifications ~\cite{DBLP:conf/cvpr/HeZL19} and model extraction attacks~\cite{DBLP:conf/asiaccs/CaoJG21, DBLP:conf/iclr/LukasZK21}. 
Cao \etal~\cite{DBLP:conf/asiaccs/CaoJG21} present an adversarial example based algorithm to generate datapoints near the decision boundary and utilize the transferability gap of those data to identify the piracy models. However, the performance of this work is not stable across different model architectures. 
Lukas \etal~\cite{DBLP:conf/iclr/LukasZK21} also adopt the transferability to craft synthesized datapoints named ``conferrable examples" that only transfer to piracy models instead of homologous models.
Crafting conferrable examples involves training up to 30 models and then backpropogating through them to obtain a gradient update. This leads to a huge overhead cost. 
\section{Problem Formulation}

\subsection{Model Definition}
Now we formally define the IP protected DNN models and the \textit{piracy models}. 
Consider a problem domain denoted by $\mathcal{X} \subset \xdomain$. Each element $\mathbf{x} \in \mathcal{X}$ is labeled by one of $N$ classes, say $i$-th class, denoted by a one-hot vector $l(\mathbf{x}) \in \ydomain$. A DNN model is a function $f : \xdomain \to \ydomain$ which takes as input $\mathbf{x} \in \xdomain$, and outputs a vector $f(\mathbf{x}) \in \ydomain$ with $i$-th entry $f(\mathbf{x})_i$ denoting the model's confidence that $\mathbf{x}$ is from the $i$-th class. 

\begin{definition}
(IP Protected DNN Model).  
A DNN model owned by a model owner $u$ is denoted by $\vmodel{u} : \xdomain \to \ydomain$. It is trained by the model owners on their dataset $\vdata{u} \subset \{ (\mathbf{x},
l(\mathbf{x})) \mid \mathbf{x} \in \mathcal{X} \}$, aiming to optimize
\begin{equation}
    \mathbb{P}_{\mathbf{x} \sim \mathcal{X}}(\argmax_{k}(\vmodel{u}(\mathbf{x})_{k}) = i \land l(\mathbf{x})_i = 1).
\end{equation}
\end{definition}

Two different model owners, say $u$ and $v$, might have slightly different training datasets, model structures, training processes, their trained models ($\vmodel{u}$ and $\vmodel{v}$) are highly similar but considered independent, we refer to as \textit{homologous models}. In contrast, in model extraction attacks, an adversary may query a victim model $\vmodel{u}$ with her chosen inputs and obtains the model outputs which can be used as labels to train a model. We refer to it as \textit{piracy model} and give a formal definition as follows.

\begin{definition}
(Piracy Model). A piracy model obtained by an attacker who launches model extraction attacks on a victim model $\vmodel{u}$ is denoted by $\pmodel{u}: \xdomain \to \ydomain$. It is trained on her dataset $\pdata{u} \subset \{ (\mathbf{x}, \vmodel{u}(\mathbf{x}))\mid\mathbf{x} \in \mathcal{X} \}$, aiming to optimize 
\begin{equation}
    \mathbb{P}_{\mathbf{x} \sim \mathcal{X}}(\argmax_{k}(\pmodel{u}(\mathbf{x})_{k}) = \argmax_{k}(\vmodel{u}(\mathbf{x})_{k}). 
\end{equation}
\end{definition}

\subsection{Threat Model}

The threat model considered in this paper involves a model owner who deploys its trained model $\vmodel{u}$ as a cloud service and an adversary who tries to launch model extraction attacks against $\vmodel{u}$ and deploys the piracy model $\pmodel{u}$ for financial benefits. The model owner here acts as both the \textit{victim} and the \textit{defender} who aims to verify whether a suspect model $\smodel{}$ is a piracy  or homologous model of $\vmodel{u}$.

\mypara{Attacker’s Capability and Knowledge.}
The attacker has black-box access to a victim model $\vmodel{u}$ and 
knows its problem domain.
To evade potential model extraction detection schemes, the adversary may also apply various modifications (\eg, fine-tuning, compression, pruning and adversarial training) to piracy models.

\mypara{Defender's Capabilities and Knowledge.}
The defender (victim) has white-box access to its model $\vmodel{u}$ (\ie, model parameters, hyper-parameters, and training dataset) and black-box access to a suspected model $\smodel{}$ with a limited number of queries. 
Specifically, the defender has no knowledge about the suspect model's architecture, parameters, hyper-parameters, nor the attacker's data used during the extraction.

\subsection{Design Overview}
\label{sec:design_goals}
In this paper, we propose to use UAPs to capture the global geometric information of a DNN model's decison boundary for model extraction detection. Universal Adversarial Perturbation suggests that a carefully selected perturbation vector $\uap \in \xdomain$ of a model $\modelsymbol{}$ can fool the  model on almost all datapoints drawn from the problem domain $\mathcal{X} \subset \xdomain$. 
Formally, a UAP $\uap$ is 
$(\xi,\delta)$-universal such that
\begin{equation}
    \begin{aligned}
        % ||\mathbf{v}||_2 &\leq \xi  \\
    \mathbb{P}_{\mathbf{x} \sim \mathcal{X}} (\argmax_{k}{\modelsymbol{}(\mathbf{x} + \uap)_k} &\ne    \argmax_{k'}{\modelsymbol{}(\mathbf{x})_{k'}}) \geq 1 - \delta, \\
    s.t., ~~ ||\mathbf{v}||_2 &\leq \xi
  \end{aligned}
\end{equation} 
where $f(\cdot)$ is the output probability vector. 
A UAP $\mathbf{v}$ can be viewed as a natural defect of the model $\modelsymbol{}$ which exposes the geometric correlations between different local gradients of the model's decision boundary. In fact, for a given model, there are a bunch of UAPs that exist in a low-dimensional subspace in which most of the normal vectors of decision boundaries lie, as pointed by Moosavi-Deafooli \etal~\cite{DBLP:conf/cvpr/Moosavi-Dezfooli17}.  

The UAP subspaces of two homologous models are independent since the decision boundaries are formed via two independent training processes. In contrast, Papernot \etal~\cite{DBLP:conf/ccs/PapernotMGJCS17} use chi-square test to statistically verify that datapoints' gradients of piracy models are dependent on those of the victim model. We observe that this dependency is maintained by UAPs. We postpone the detail of this observation to \autoref{sec:observation_exp} and continue our design overview. 
Since calculating a UAP requires white-box access to the model, it is infeasible for the defender to obtain the UAPs of suspect model $\smodel{}$ and compare their similarity with $\vmodel{u}$. Alternatively, with white-box access to the victim model $\vmodel{u}$, the defender (victim) could generate a UAP $\uap$ and verify whether $\uap$ lies in the UAP subspace of the suspect model $\smodel{}$. Specifically, we propose the following two primitives for the verification:

\mypara{Fingerprint Generation.}
To verify whether a vector $\uap$ lies in the UAP subspace of a model $\modelsymbol{}$, we propose to design a fingerprints generation function $\fpfunc$ which captures the \textit{fingerprints} of how the model $\modelsymbol{}$ behaves around $\fplen$ datapoints $\mathbf{x}_1,\dots, \mathbf{x}_\fplen$ with regards to $\uap$, denoted by $\fpfunc(\modelsymbol{}, \uap, [\mathbf{x}_1,\dots, \mathbf{x}_\fplen])$.

\mypara{Fingerprint Verification.}
With fingerprints of both the victim model and the suspect model, the defender needs to determine whether the suspect model is a piracy model or a homologous model. Particularly, we propose to design an encoder ${E_\theta}$ with parameters $\theta$ to map the fingerprints of models to latent space, such that the mapped fingerprints of a victim model and its piracy model have a large similarity (\eg, cosine similarity) while the mapped fingerprints of a victim model and a homologous model has a small similarity. Particularly, the encoder aims to optimize: 
\begin{equation}
    \resizebox{.9\hsize}{!}{ $
    \begin{aligned}
        \max \limits_{\theta} \; \mathbb{E}(sim(\vmodel{u}, & \pmodel{u})) \, - 
    \mathbb{E}(sim(\vmodel{u}, \vmodel{v})), \\
    \text{where}~~sim(\modelsymbol{}_a, \modelsymbol{}_b) &= cosine(E_\theta(\mathcal{F}_a), E_\theta(\mathcal{F}_b)) \\
    \mathcal{F}_a &= \fpfunc(\modelsymbol{}_a, \uap, [\mathbf{x}_1,\dots, \mathbf{x}_\fplen]) \\
    \mathcal{F}_b &= \fpfunc(\modelsymbol{}_b, \uap, [\mathbf{x}_1,\dots, \mathbf{x}_\fplen]).
    \end{aligned}
    $}
\end{equation}
\vspace{-4mm}

\section{UAP based Fingerprinting}

In this section, we first explain the observation on which our design is based. Then we introduce the design of fingerprint generation and fingerprint verification.

\subsection{Observation Explanation}
We first show the relation among UAP subspaces of the victim model, homologous models, and piracy models. 

\mypara{\textsc{Observation}:}
\textit{A victim model's UAP subspace is consistent with its piracy model's UAP subspace and is inconsistent with a homologous model's UAP subspace.} 
\label{sec:observation_exp}

\begin{figure}[t]
    \centering
	\begin{subfigure}[t]{0.22\textwidth}
	    \centering
    	\includegraphics[width=\linewidth, height=0.702\linewidth]{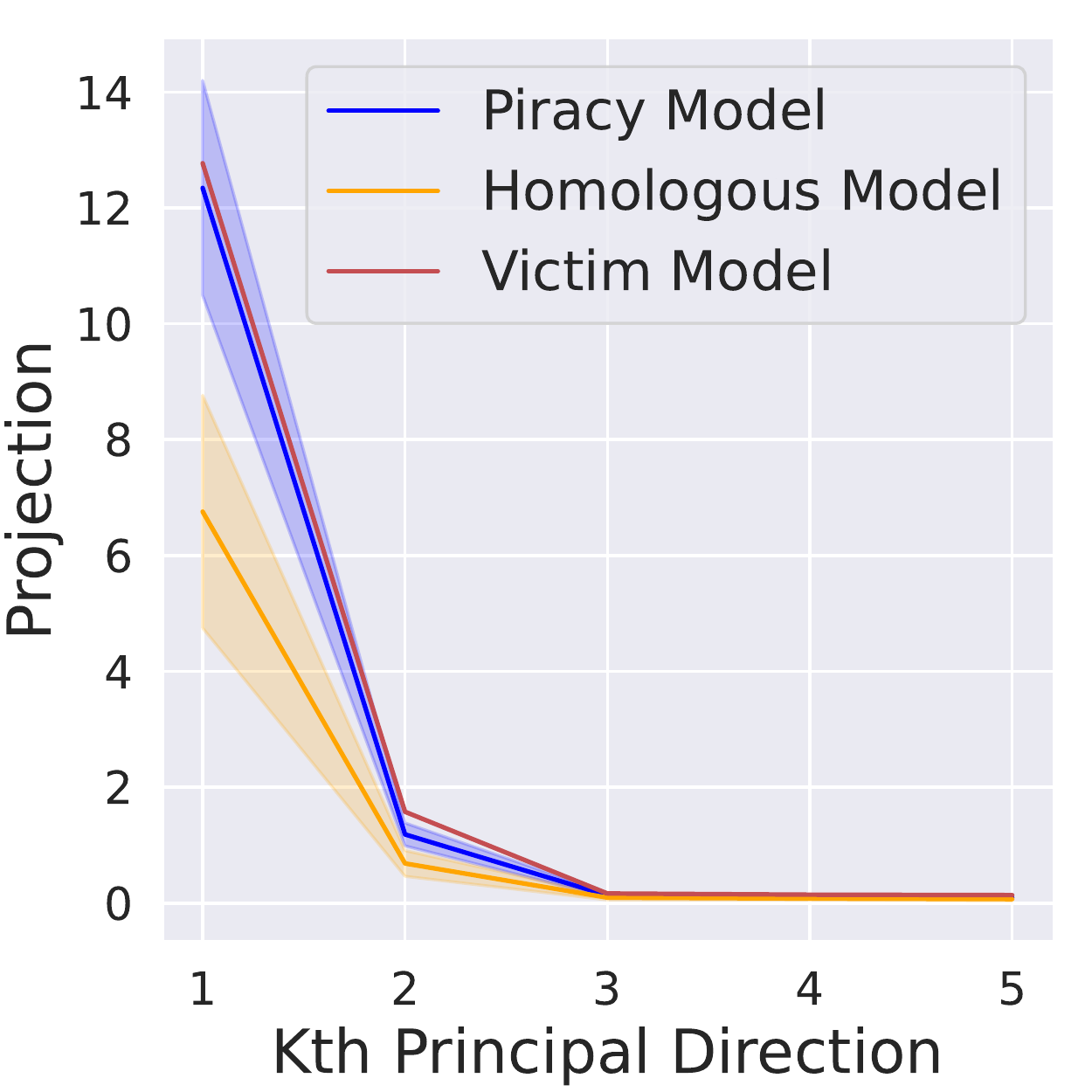}
    	\caption{SVD on DNN models' UAPs.}
    	\label{fig:poc:svd_bar}
    \end{subfigure}	
    \hfill
    \begin{subfigure}[t]{0.21\textwidth}
		\centering
		\includegraphics[width=\linewidth]{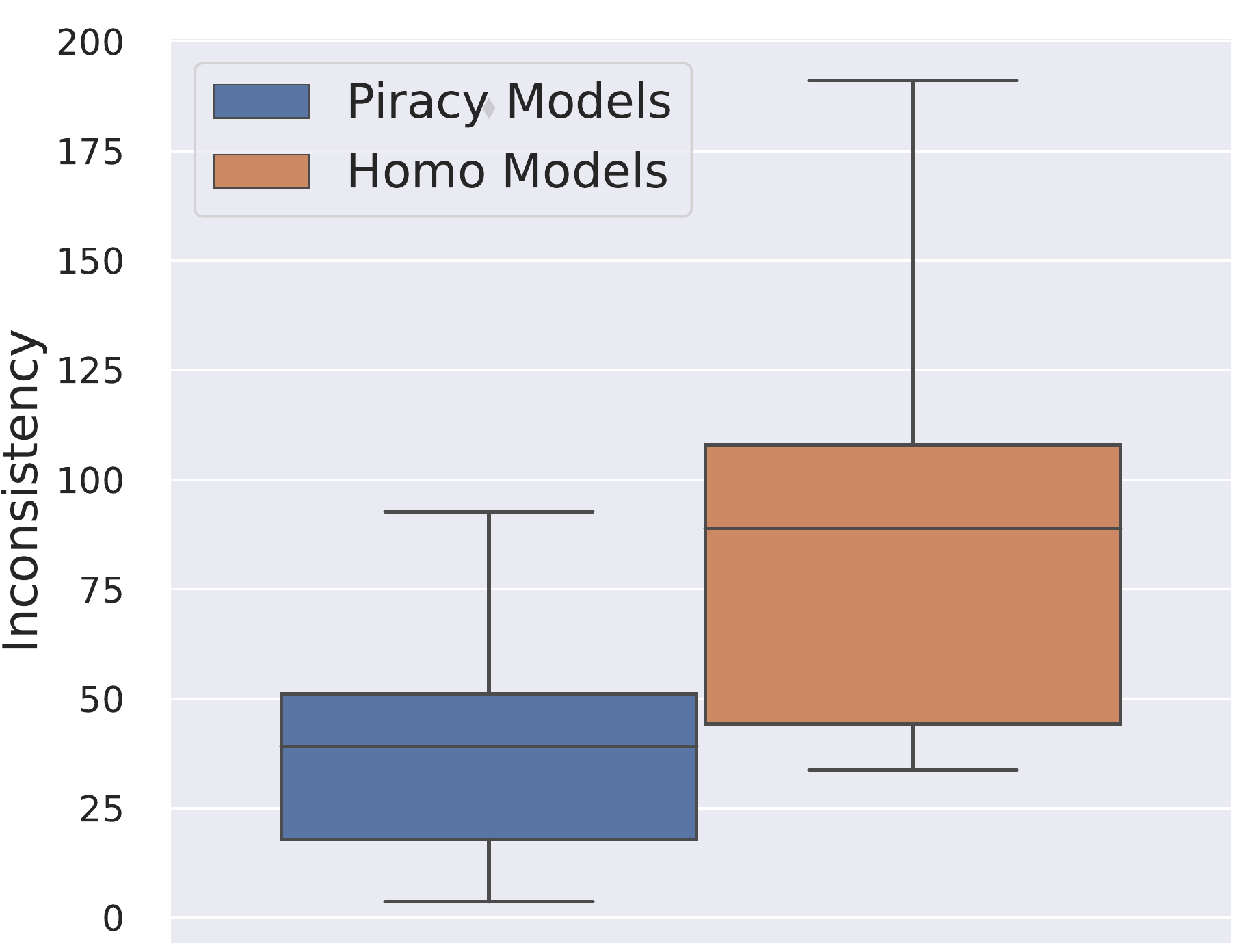}
		\caption{Inconsistency distributions.}
		\label{fig:poc:svd_inconsis}
	\end{subfigure}
	\vspace{-2mm}
    \caption{\small (a) Projections of $\vmodel{u}$, piracy and homologous models on top-5 principal directions of $\vmodel{u}$, the dark line indicates mean value over 20 models and light intervals indicate STD; (b) Inconsistency distribution of piracy models and homologous models calculated according to \autoref{eq:sec4:poc_incssts}.}
	\label{fig:poc:svd}
	\vspace{-3mm}
\end{figure}

We model the dependency of a suspect model $\smodel{}$ on a victim model $\vmodel{u}$ as their \textit{consistency}, which is defined as $\ell_2$ distance between the projections of these two models' UAPs on a set of orthogonal basis formed by principal directions of $\vmodel{u}$'s UAP matrix. To obtain this basis, we perform singular-value decomposition on $\vmodel{u}$'s UAP matrix and choose its right singular vector basis.

Let $V_{\smodel{}}=\{ \mathbf{v}_{\smodel{}}^1, \dotsc, \mathbf{v}_{
\smodel{}}^L \}$ be UAPs of the suspect model $\smodel{}$. $V_{\vmodel{u}}=\{ \mathbf{v}_{\vmodel{u}}^1, \dotsc, \mathbf{v}_{
\vmodel{u}}^L \}$ be UAPs of the victim model. 
After performing SVD on $V_{\vmodel{u}}$, there is a $r$-dimensional orthogonal basis $\{ v_1, v_2, \cdots, v_r \}$, where $r$ is the rank of $V_{\vmodel{u}}$.
We define the distribution inconsistency of UAPs between $V_{\smodel{}}$ and $V_{\vmodel{u}}$ as $Inconsist_{\vmodel{u}}(\smodel{})$:
\begin{equation}
\label{eq:sec4:poc_incssts}
\resizebox{.9\hsize}{!}{ $
    \sum_{0 \leq m \leq r} \big( {\sum_{0 \leq i \le
    L} (\mathbf{v}_{\smodel{}}^i \cdot v_m)^2 - \sum_{0 \leq j \leq L} (\mathbf{v}_{\vmodel{u}}^i \cdot v_m)^2} \big) ^ 2.
$}
\end{equation} 
We generate $m$ (input dimension) UAPs for each model on FMNIST dataset to form a square UAP matrix with its rank equals $m$. \autoref{fig:poc:svd_bar} shows that the piracy models' UAPs has similar projections as the victim, whereas the homologous models loosely follow principal directions. \autoref{fig:poc:svd_inconsis} shows that $Inconsist_{\vmodel{u}}(\pmodel{u})$ is 3 times smaller than $Inconsist_{\vmodel{u}}(\vmodel{v})$ on FMNIST dataset. We conclude that UAPs of piracy and homologous models differ and can be used to differentiate these two types of models. 

\begin{figure}[t]
    \centering
	\begin{subfigure}[b]{0.23\textwidth}
    	\includegraphics[width=\linewidth]{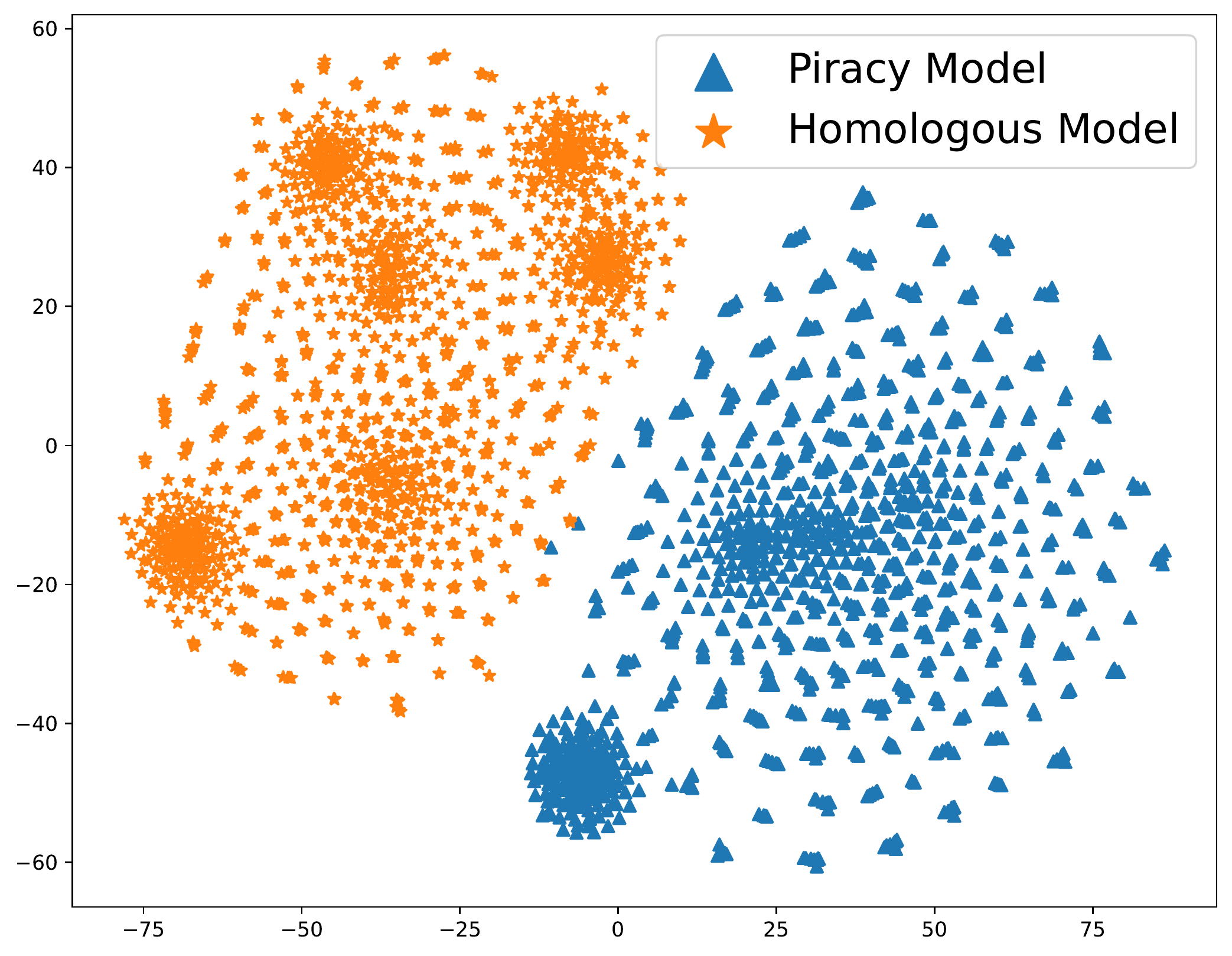}
    	\label{fig:poc:uap_figer}
    \end{subfigure}	
    \hfill
    \begin{subfigure}[b]{0.23\textwidth}
		\centering
		\includegraphics[width=\linewidth]{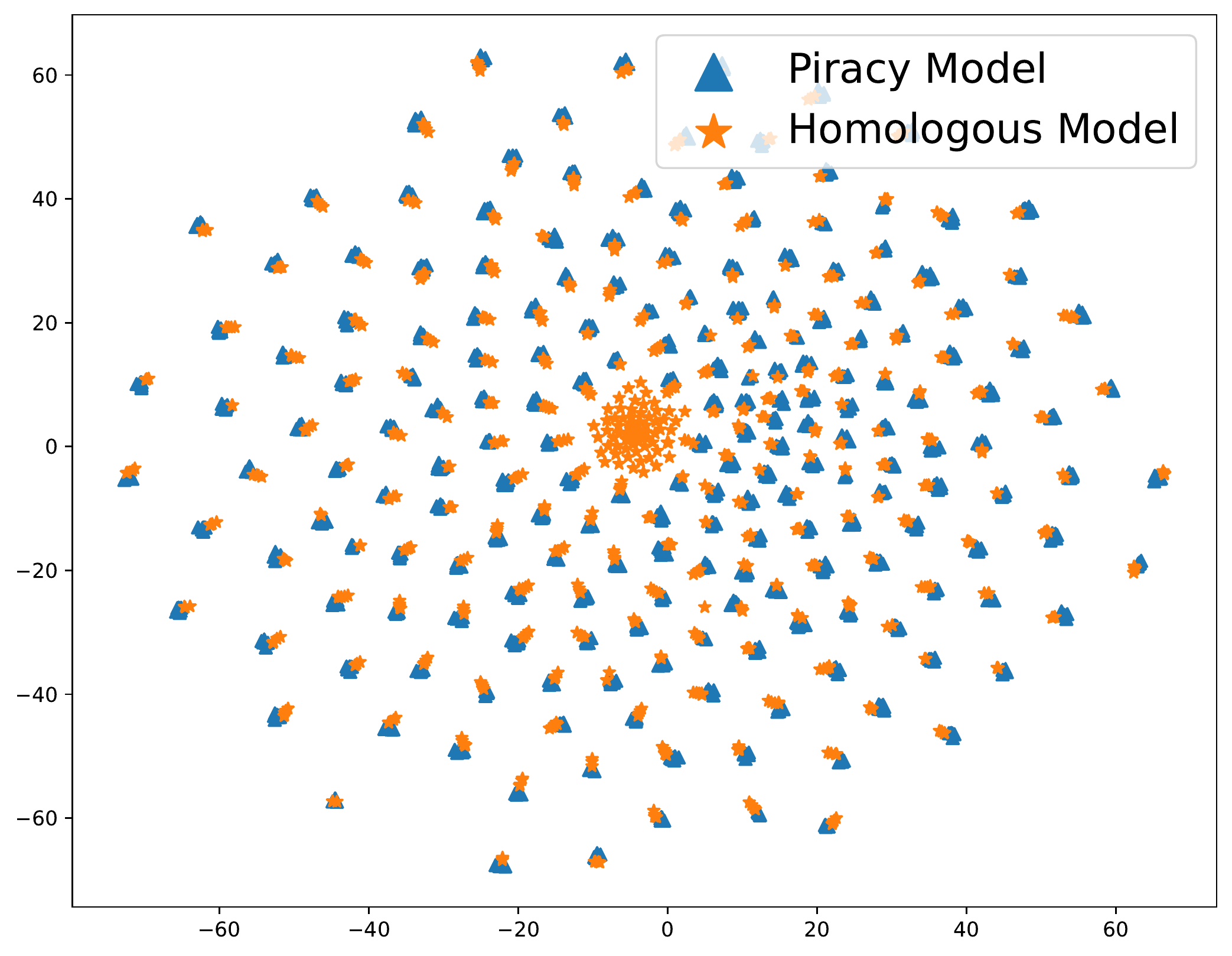}
		\label{fig:poc:ae_figer}
	\end{subfigure}
	\vspace{-5mm}
    \caption{\small $t$-SNE visualization of fingerprints. UAP-based fingerprints (left) are naturally distinguishable compared with local adversarial perturbation based fingerprints (right). (FMNIST)}
	\label{fig:poc:uap_ae}
	\vspace{-3mm}
\end{figure}

\subsection{Fingerprint Generation} \label{sec4:quptse}
We now define a fingerprint generation function as described in \autoref{sec:design_goals} as follows:
\begin{equation} 
\label{eq:sec4:fgprt}
\begin{aligned}
    &\fpfunc(\modelsymbol{}, \uap, (\mathbf{x}_1, \cdots, \mathbf{x}_\fplen)) \\= &  [\modelsymbol{}( \mathbf{x}_1) , \modelsymbol{}(\mathbf{x}_1 + \uap) , \cdots , 
    \modelsymbol{}(\mathbf{x}_\fplen) , \modelsymbol{}(\mathbf{x}_\fplen + \uap)].
\end{aligned}
\end{equation}
The goal of $\fpfunc$ is to capture how the given model's outputs change around the given datapoints before and after adding the UAP. Intuitively, if $\uap$ is a UAP of $\smodel{}$, adding $\uap$ to samples will significantly reduce the confidence of $\smodel{}$ on its original predict class. Otherwise, adding $\uap$ will not push samples to a decision boundary.

The next step is to select $\fplen$ datapoints that could better profile the decision boundary. Besides using more datapoints (\ie, larger $\fplen$), we would like the datapoints spread uniformly throughout entire decision boundaries to capture diverse information. Thus, we perform K-means to cluster all training datapoints of the victim model into $\fplen$ clusters according to their representation vector in the last layer of the model, and select one datapoint from each cluster.

We compare the effectiveness of UAP based fingerprints to local adversarial perturbation(LAP) based fingerprints of three types of models (the victim, piracy and homologous). As shown in \autoref{fig:poc:uap_ae}, for UAP based fingerprints, models with the same type form one cluster, which is distant from the clusters of models with different types. In contrast, LAP based fingerprints of models with different types mix together and are indistinguishable. 
Please be referred to {\color{magenta}supplementary materials} for more discussions.

\subsection{Fingerprint Verification}
\label{sec:encoder_training}

\begin{figure}[t]
    \centering
    \includegraphics[width= 0.9\linewidth]{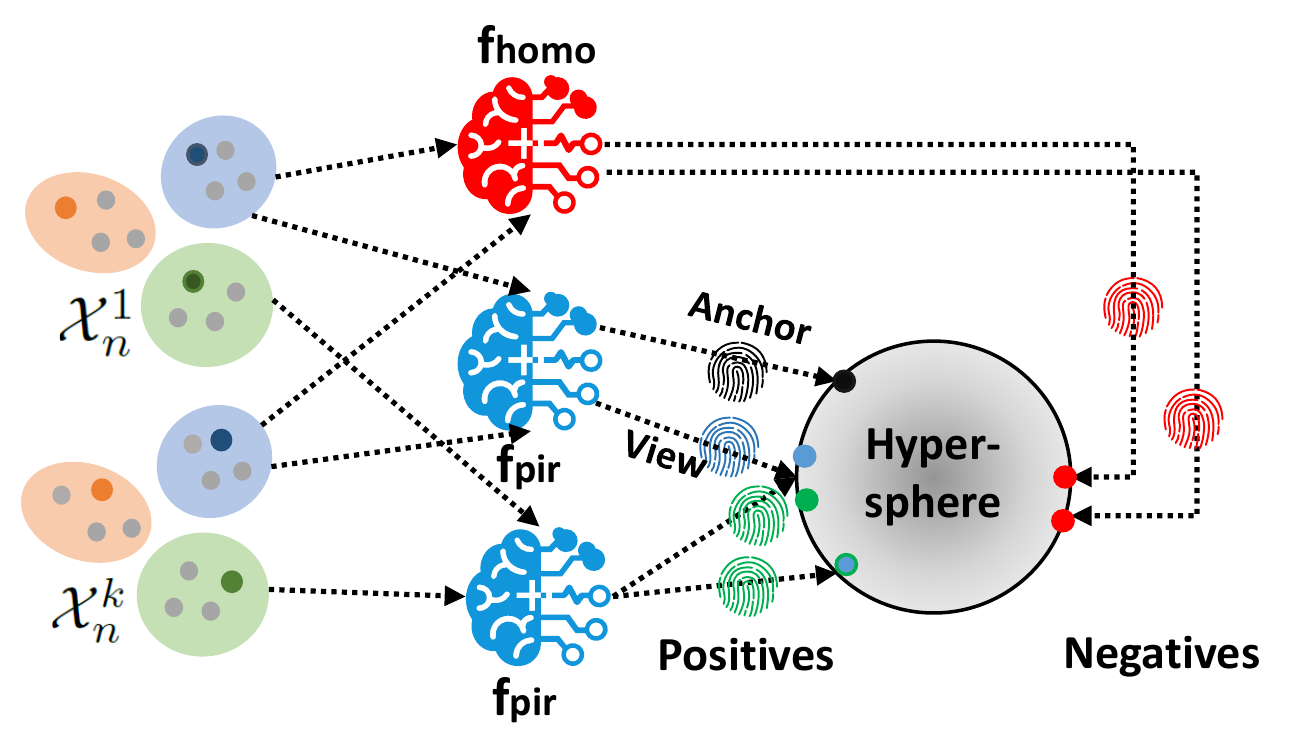}
    \vspace{-3mm}
     \caption{\small Illustration of contrastive learning. $\mathcal{X}_n^1 \cdots \mathcal{X}_n^k$ are k sets of datapoints used to create ``views" (blue fingerprint is an augmented view for black one) . Piracy fingerprints are positive with each other (green) and negative with homologous fingerprints (red). Encoder projects fingerprints to a hyper-sphere. }
   	\label{fig:fw:cl}
   	\vspace{-3mm}
\end{figure}

We leverage an encoder to learn knowledge contained in fingerprints and output a human-comprehensible similarity score. 
The encoder projects the features of fingerprints into a latent space and one can easily compares two fingerprints by their representations' cosine similarity in the space.

Simply training an encoder (\eg, AutoEncoder) can only extracts common features of piracy fingerprints and other fingerprints that have different features will not be mapped near to embedding space. As homologous models are highly similar to the victim model, it fails to project them away from piracy one. We leverage the supervised contrastive learning~\cite{DBLP:conf/nips/KhoslaTWSTIMLK20} to emphasize such differentiation on homologous models.

Precisely, we assign label 0 to victim and piracy fingerprints, label 1 to homologous fingerprints. As demonstrated in \autoref{fig:fw:cl}, the encoder projects positive pairs onto the same part on the hyper-sphere (left side) and projects negative pairs onto the opposite part (right side). For self-supervised contrastive learning, a positive pair $(\mathbf{x}, \mathbf{\tilde{x}})$ refers to an input $\mathbf{x}$ and its view $\mathbf{\tilde{x}}$, all other inputs and their views are negative to $\mathbf{x}$. For supervised contrastive learning, the positive pairs are all inputs having the same label as $\mathbf{x}$ and their views, negative pairs are all other inputs and their views.    

To generate positive pairs for a given fingerprint in contrastive learning, we propose a novel data augmentation strategy as follows.

\mypara{Multi-views Fingerprints Augmentation.}
We denote $\mathcal{X}_n$ is the $n$ datapoints selected from $n$ different clusters. For each datapoint $\mathbf{x}_i$ in $\mathcal{X}_n$, we choose its $k$ nearest neighbors to form $\mathcal{K}_i$ according to their representations in the output layer. 
We perform sampling without replacement in each cluster $\mathcal{K}_i$ and obtain $k$ sets of datapoints, denoted by $\mathcal{X}_n^1, \cdots, \mathcal{X}_n^k$ (See \autoref{fig:fw:cl} for an illustration).
In this way, we further generate $k$ positive views $\{\mathcal{F}(f, \mathbf{v}, \mathcal{X}_n^1), \cdots, \mathcal{F}(f, \mathbf{v}, \mathcal{X}_n^k)\}$ for the given fingerprint $\mathcal{F}(f, \mathbf{v}, \mathcal{X}_n)$. Please see {\color{magenta}supplementary materials} for more details, including the evidence that positive views are the most similar fingerprints among others.

\mypara{Supervised Contrastive Loss.}
We now describe our supervised contrastive loss as follows. Within a multiviewed batch, let $i \in I \equiv \{1, ..., k_N\}$ be batch index and $C(i) = I \backslash \{i\}$. Let $\Psi(i) := \{ \mu \in C(i) | \tilde{y_{\mu}} = y_i \}$ be indexes of positive pairs for the $i$-th sample. 
Then our supervised contrastive loss is:
\begin{equation} \label{eq:spd_cts_loss}
  \mathcal{L} = \sum_{i \in I} -\frac{1}{|\Psi(i)|} \sum_{\mu \in \Psi(i)} \log \frac{ e ^ {{sim(\mathbf{z}_i, \mathbf{z}_{\mu})} / \tau}} {  \sum_{ \nu \in C(i) } e^{sim(\mathbf{z}_i, \mathbf{z}_{\nu}) / \tau } }, 
\end{equation}
where $N$ is the size of mini-batch.
The encoder consists of two parallel networks, which is similar to SimCLR~\cite{DBLP:conf/icml/ChenK0H20}
The overall procedure of model extraction detection is represented by \autoref{algo:framework}.

\begin{algorithm}[t]\footnotesize
\KwIn{
suspect model $\smodel{}$, 
victim's model $\vmodel{u}$, it's UAP $\mathbf{v}$ and training data $\mathcal{D}$, 
number of clusters $n$, 
number of fingerprints views $k$, 
a set of piracy models $\Phi$ and homologous models $\Upsilon$,
batch size $N$ and loss function $\mathcal{L}$ for constrastive learning in \autoref{eq:spd_cts_loss}.
}
\KwOut{Trained encoder $E_{\theta}$, similarity $\mathbf{s}$ between $\smodel{}$ and $\vmodel{u}$.}

\tcc{ Fingerprints Generation}
\DontPrintSemicolon
\SetKwFunction{FMain}{$\mathcal{F}$}
\SetKwProg{Fn}{Function}{:}{}
\Fn{\FMain{$f, \{ \mathbf{x}_1, \cdots, \mathbf{x}_n \}, \mathbf{v}  $}}{
    $\mathcal{X} \leftarrow \{\}$ \;
    \For{$i\in \{1,..,n\}$}{
        $t_i = f(\mathbf{x}_i) \oplus f(\mathbf{x}_i + \mathbf{v}) $ \;
        $\mathcal{X} = \mathcal{X} \cup \{ t_i \}$ \;
    }
    \KwRet $\mathcal{X}$
}

\tcc{Preparing trainset for encoder $E$}
$B \leftarrow \{\}$; $\;$ $\mathcal{M} \leftarrow \{ \vmodel{u}  \} \cup \Phi \cup \Upsilon $\;
$\{C_1, C_2, \cdots, C_n \} = \;$  K-Means$(f_{\vmodel{u}}, \mathcal{D}, n) $\;

\For{$f \in \mathcal{M}$}{
    \tcc{Sample one point from each cluster without replacement}
    \For{$i \in \{1,...,k\}$}{
        $\{ \mathbf{x}_1, \cdots, \mathbf{x}_n \}^i \leftarrow C_1 \times C_2 \cdots \times C_n$ \;
        $B = B \cup {\mathcal{F}(f, \{ \mathbf{x}_1, \cdots, \mathbf{x}_n \}^i, \mathbf{v})}$
    }
}

\tcc{Training by contrastive loss}
Initial parameters $\theta$ of the encoder $E$\;
$E_{\theta} \leftarrow $Training($B, \mathcal{L}$)\;

\tcc{Verifying suspect model}
$\mathbf{s} = cosine(E_{\theta}(\smodel{}), E_{\theta}(\vmodel{u}))$ \;
\Return{$\mathbf{s}$}
\caption{Ownership Verification.}
\label{algo:framework}
\end{algorithm}

\section{Experiments}

\begin{figure*}[t]
    \centering
	\begin{subfigure}[b]{0.33\textwidth}
		\includegraphics[width=\linewidth]{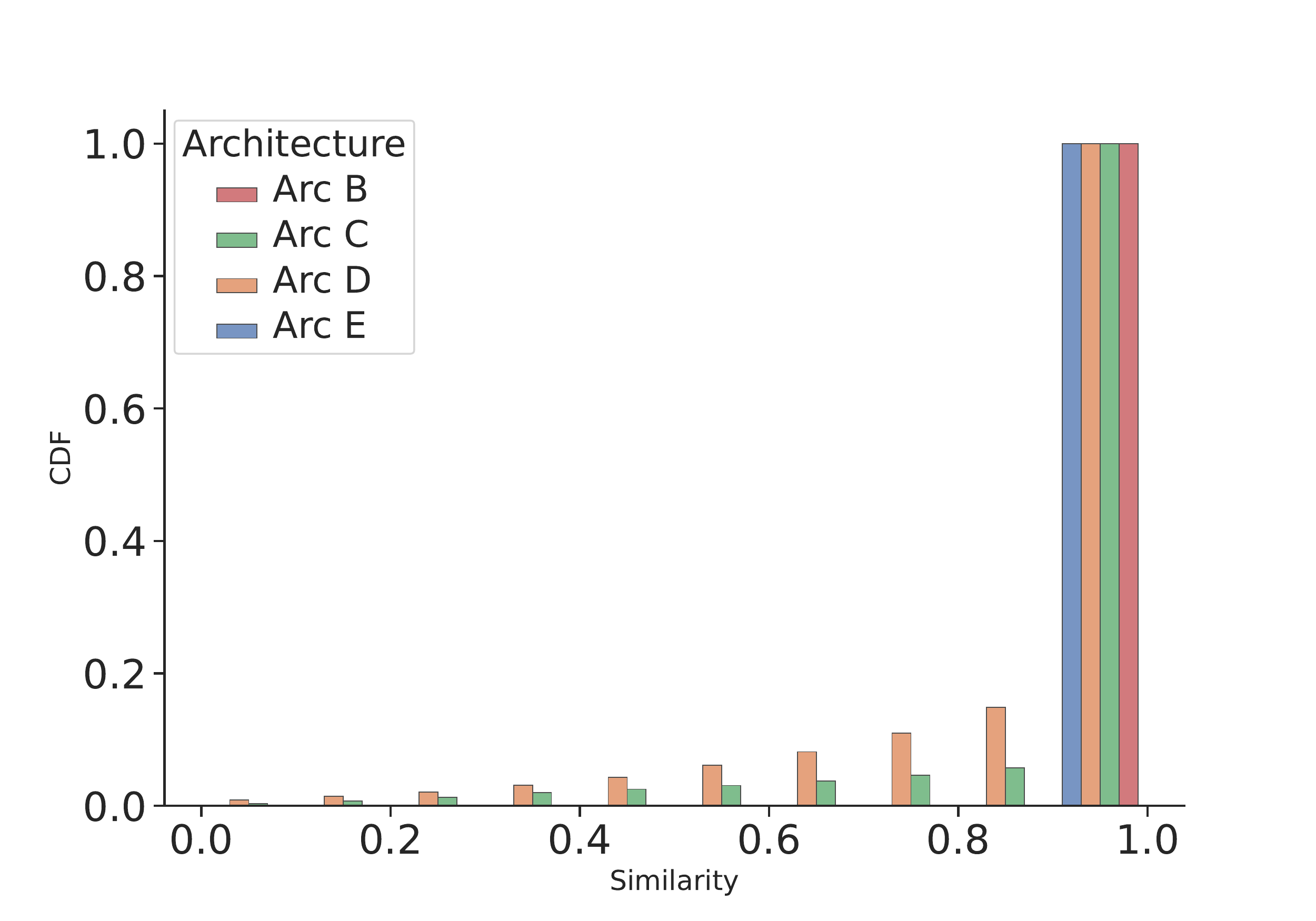}
		\caption{Sim CDF between $\vmodel{u}$ and $\pmodel{u}$(FMNIST)}
		\label{fig:eval:cdf_sim_pir_fmnist}
	\end{subfigure}
	\hfill
	\begin{subfigure}[b]{0.33\textwidth}
		\includegraphics[width=\linewidth]{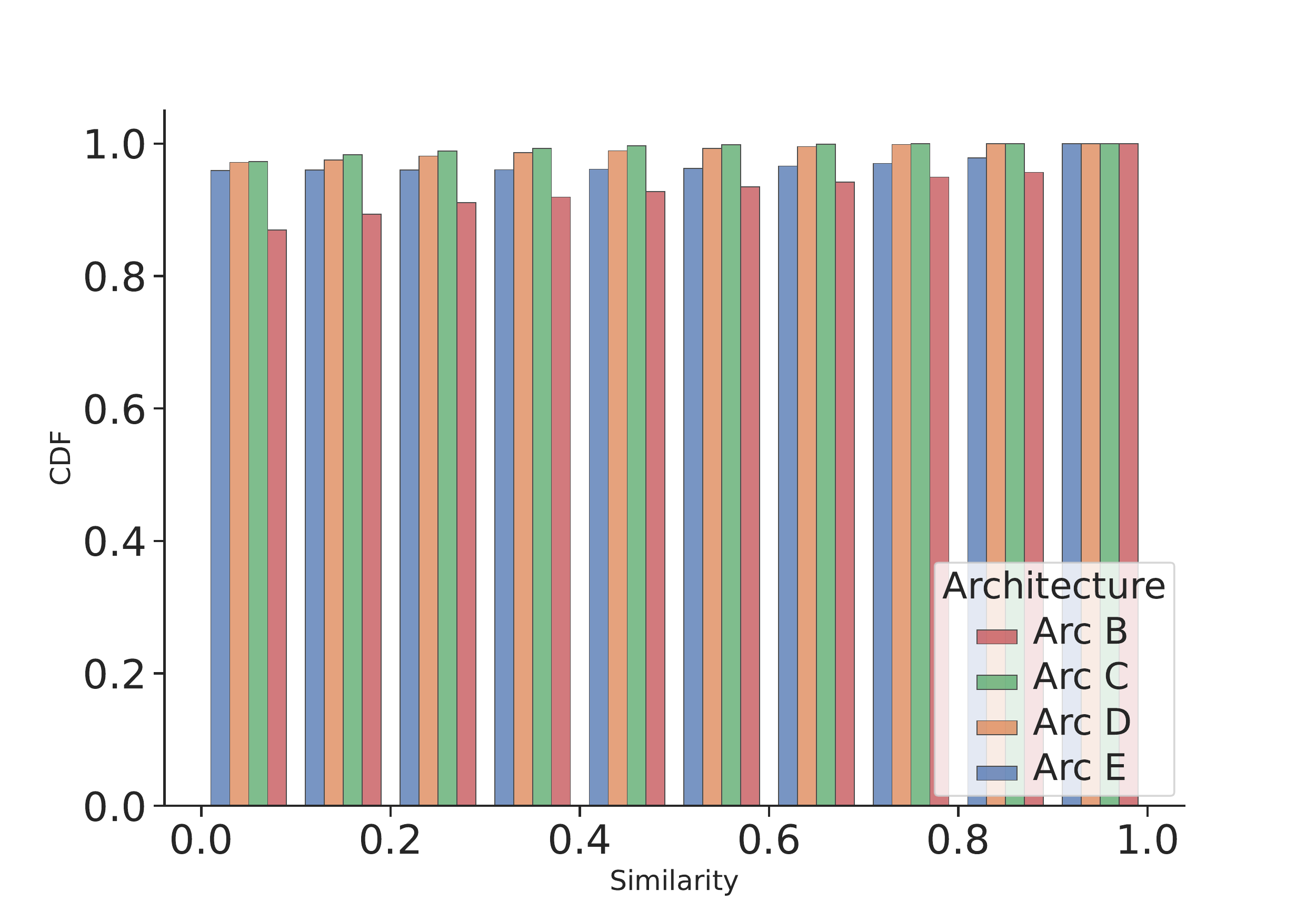}
		\caption{Sim CDF between $\vmodel{u}$and $\vmodel{v}$(FMNIST)}
		\label{fig:eval:cdf_sim_homo_fmnist}
	\end{subfigure}
	\hfill
	\begin{subfigure}[b]{0.33\textwidth}
		\centering
		\includegraphics[width=\linewidth]{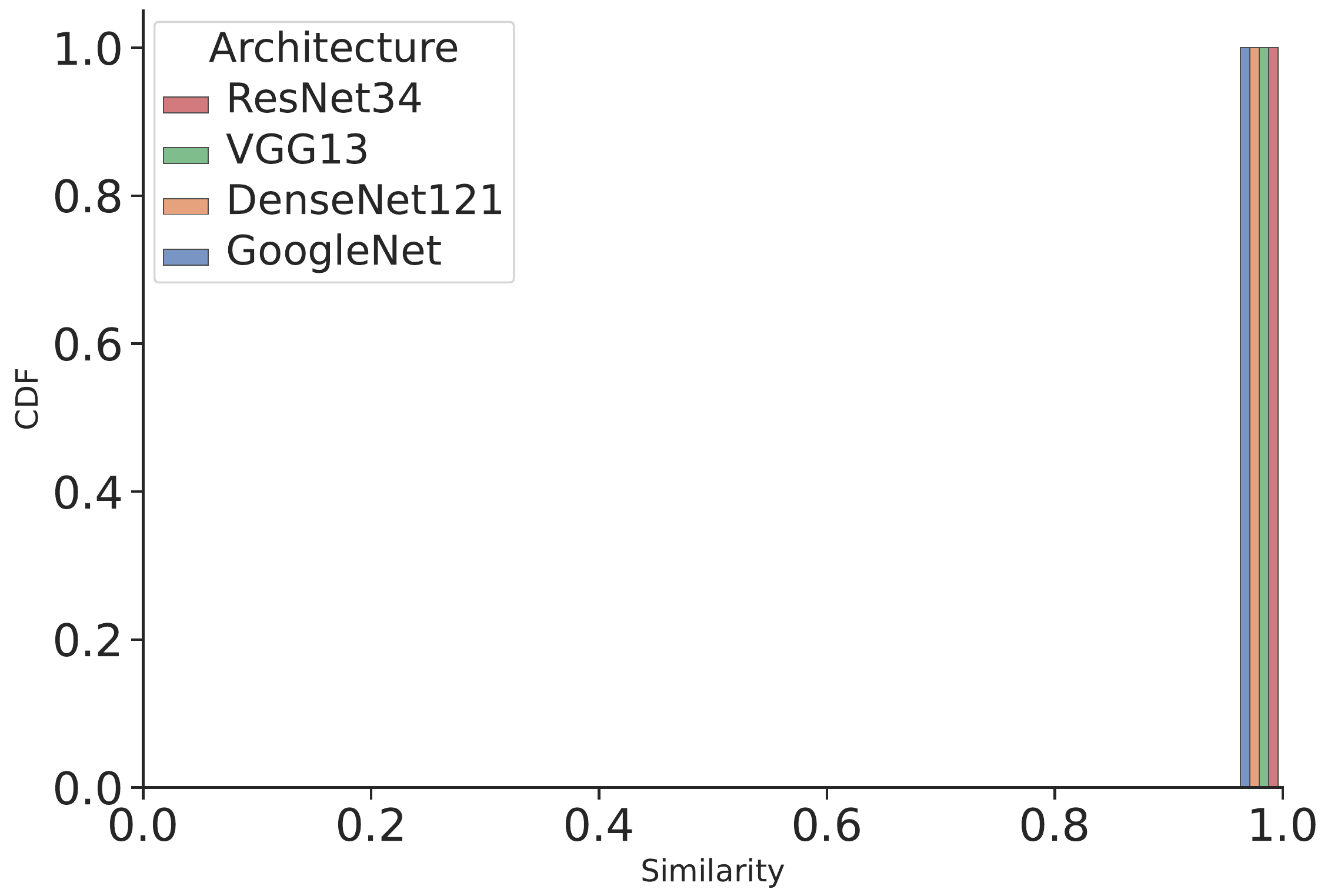}
		\caption{Sim CDF between $\vmodel{u}$and $\pmodel{u}$(CIFAR10)}
		\label{fig:eval:cdf_sim_pir_cifar10}
	\end{subfigure}\hspace{\fill}
    % \hfill
	\begin{subfigure}[b]{0.33\textwidth}
		\includegraphics[width=\linewidth]{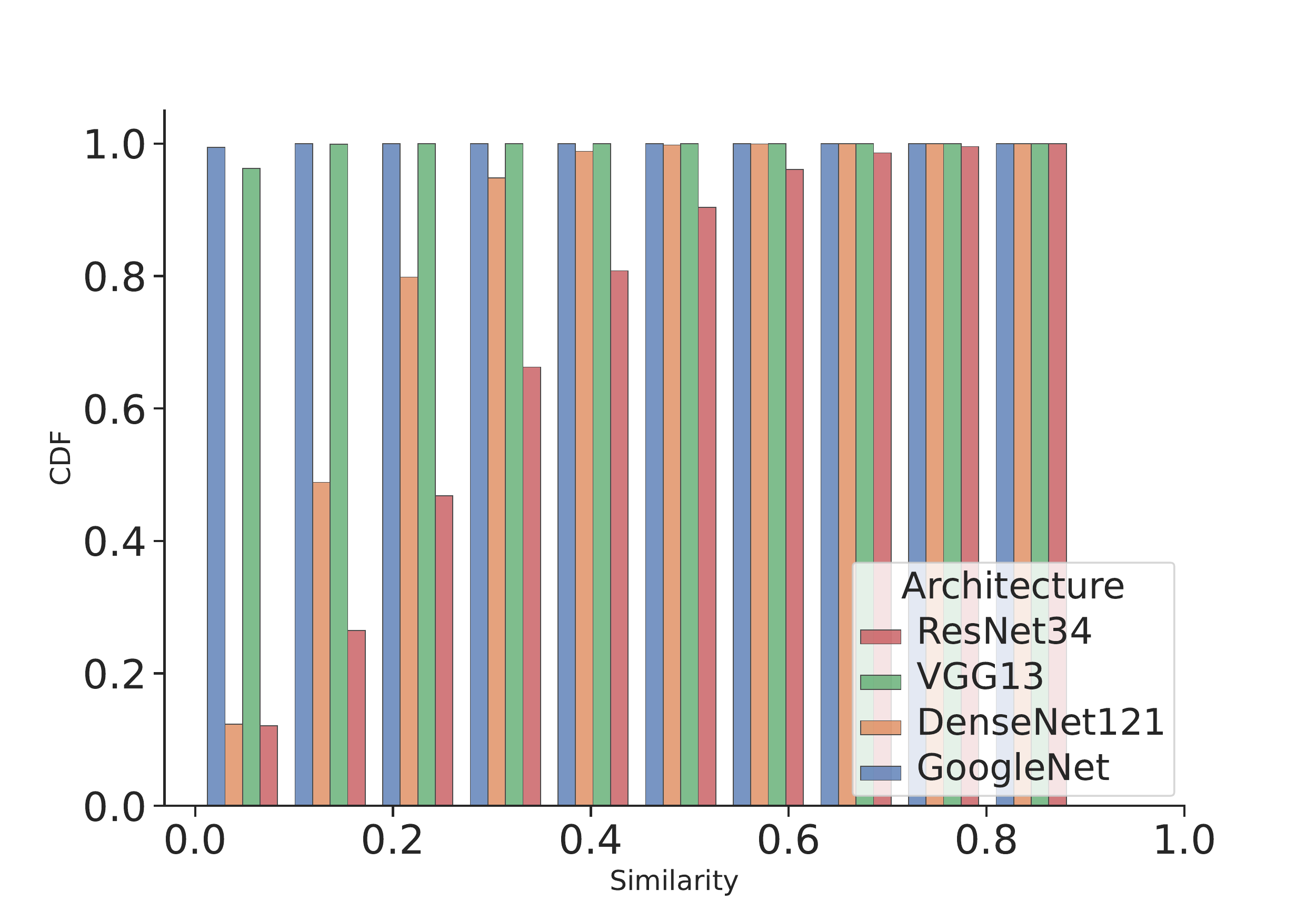}
		\caption{Sim CDF between $\vmodel{u}$and $\vmodel{v}$(CIFAR10)}
		\label{fig:eval:cdf_sim_ind_cifar10}
	\end{subfigure}
	\begin{subfigure}[b]{0.33\textwidth}
		\includegraphics[width=\linewidth]{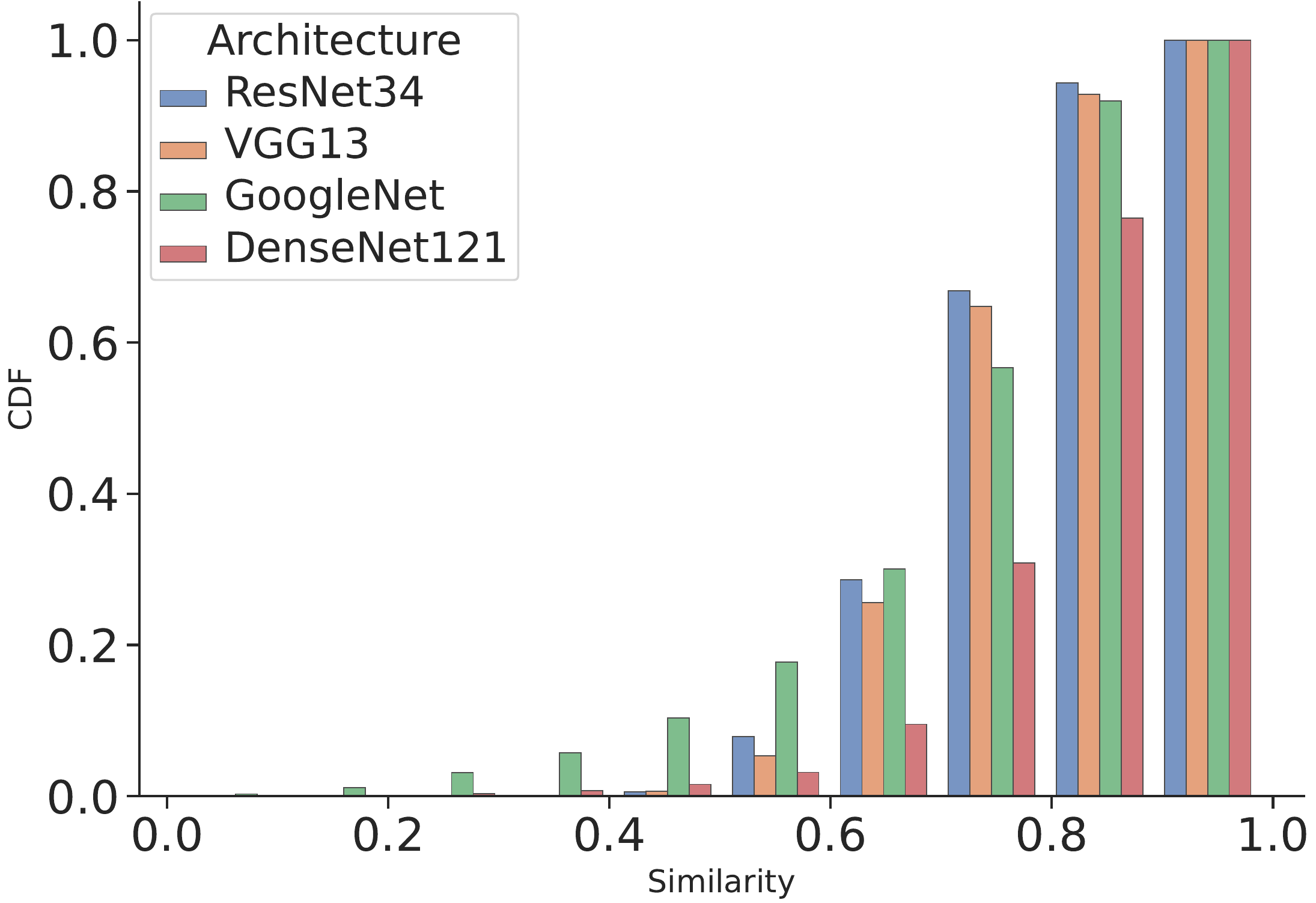}
		\caption{Sim CDF between $\vmodel{u}$ and $\pmodel{u}$(T-ImageNet)}
		\label{fig:eval:cdf_sim_ind_ti}
	\end{subfigure}
	\hfill
	\begin{subfigure}[b]{0.33\textwidth}
		\includegraphics[width=\linewidth]{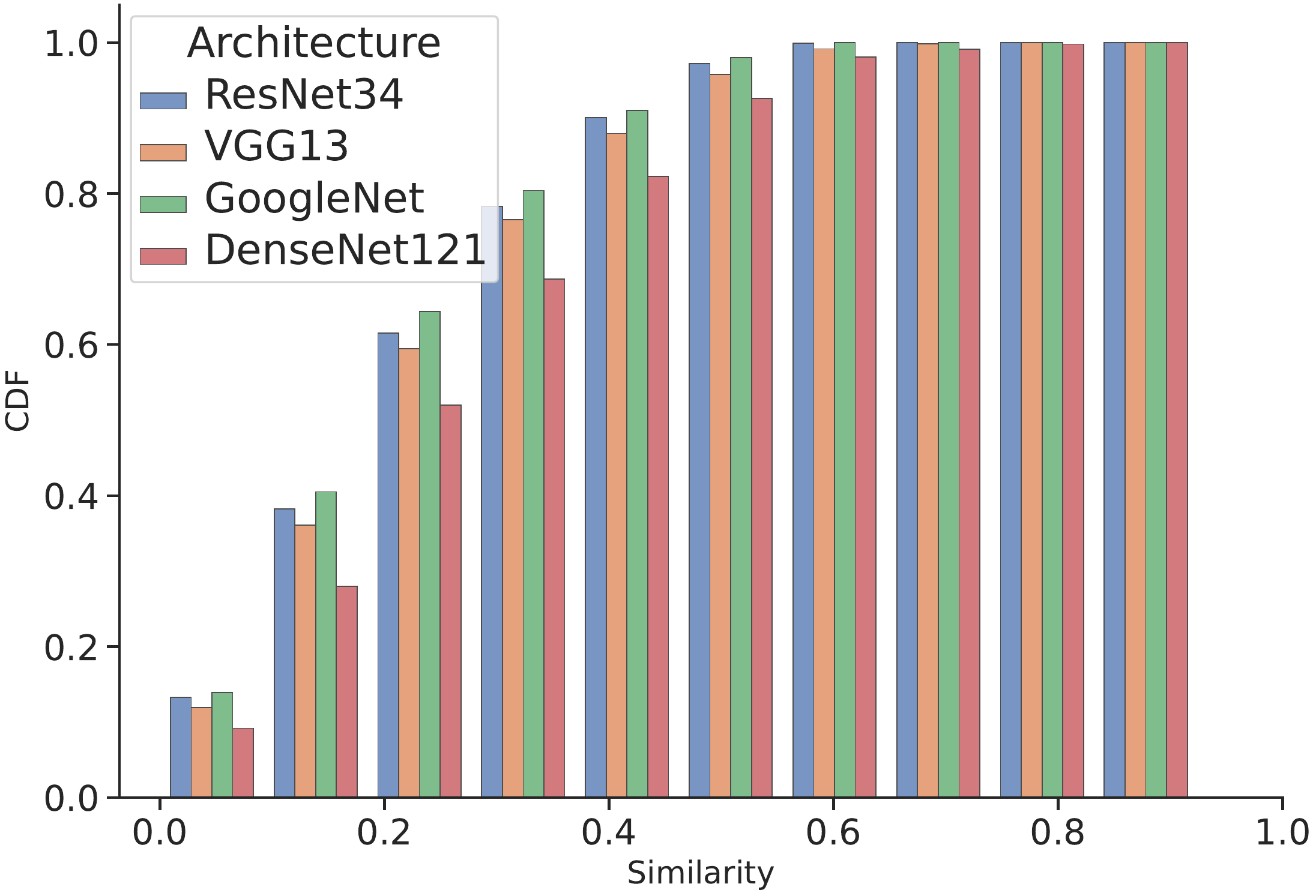}
		\caption{Sim CDF between $\vmodel{u}$ and $\vmodel{v}$ (T-ImageNet).}
		\label{fig:eval:cdf_sim_ind_ti}
	\end{subfigure}\hspace{\fill}
	\vspace{-5mm}
	\caption{\small Cumulative distribution function (CDF) of similarities between fingerprints of $f_{\mathcal{V}}$ and suspect models. For $x$ in X-axis, the Y-axis is the percentage of fingerprints that has similarity smaller than $x$. The derivation of CDF is the probability density function.}
	\label{fig:eval:cdf}
	\vspace{-3mm}
\end{figure*}

\subsection{Setup}
\mypara{Datasets.}
We evaluate our approach on three popular image classification datasets: FashionMNIST (FMNIST)~\cite{xiao2017/online}, CIFAR-10~\cite{krizhevsky2009learning}, and TinyImageNet~\cite{tiny_imagenet}.

\mypara{Model Architectures.}
For FMNIST, SOTA classification accuracy can be achieved using simple CNN models. To ensure the diversity of models, we vary attributes such as kernel size, number of layers, activation functions, dropout, training batchsize and optimizer. Details are shown in \autoref{tab:evl:arch_FMNIST}. We regroup these attributes into 5 different model architectures. The numbers inside brackets indicate the attributes is assigned to which architectures. Attributes without numbers are randomly selected during each training process. 
For CIFAR-10 and TinyImageNet, 
we evaluate our encoder in 5 different architectures: ResNet18 and ResNet34~\cite{he2016deep}, VGG16~\cite{vgg}, DenseNet121~\cite{densenet}, GoogLeNet~\cite{szegedy2015going}. 
\begin{table}[t] 
\centering
\caption{\small FashionMNIST classifier components.}
\vspace{-3mm}
\resizebox{.9\hsize}{!}{%
\begin{tabular}{lll}
\hline
                              & Attribute     & Value              \\ \hline
\multirow{4}{*}{Architecture} & Activation    & ReLU[3,5], PReLU[4], ELU[1,2]     \\
                              & Dropout       & Yes[5], No[1,2,3,4]             \\
                              & Conv ker size & 3[1,3,4], 5[2,5]                \\
                              & \#Conv layers & 2[1,3,5], 3[2], 4[4]              \\ \hline
\multirow{2}{*}{Optimziation} & Algorithm     & SGD, ADAM, RMSprop \\
                              & Batch size    & 64,128,256         \\ \hline
\end{tabular}
}
\vspace{-3mm}
\label{tab:evl:arch_FMNIST} 
\end{table}

\mypara{Model Preparation.}
For each dataset, we only assign half of the train set $D$ as the victim model's trainset, denoted as $D_v$. Piracy models are generated by following the extraction attack in ~\cite{DBLP:conf/ndss/YuYZTHJ20}. 
All generated piracy models recover $85\%, 83\%, 40\%$ performance of $\vmodel{v}$ for FMNIST, CIFAR10 and TinyImageNet, respectively. Homologous models are trained on $D_{homo}$ which is sampled from $D$ and has equal size as $D_v$, overlapping with it.

To avoid contingency, we generate 10 models for each type (piracy or homologous) and architecture (except for architecture reserved for the victim model). In total we train $81$ DNN models for CIFAR-10 and TinyImageNet respectively. We take both 4 model architectures and 3 optimizers into account and generate $241$ models. All models achieve SOTA performance.
The UAP used in this work achieves $ > 80\% $ attack success rate.
Please see more results and analyses in {\color{magenta}supplementary materials}.

\mypara{Encoder Training.}
Excluding the architecture of $\vmodel{v}$, we use the remaining architectures to train other models. We use 5 piracy models and 5 homologous models to train the encoder. 
The rest of the models are used to test the encoder. 
For three datasets, each fingerprint is consists of $100$ datapoints and we generate $200$ views for each fingerprint. For encoder training, we adopt a large batch size equals $512$ as recommended in the contrastive learning~\cite{DBLP:conf/icml/ChenK0H20}.
Note that, in experiments, to demonstrate the generalizability and robustness of our approach, we generate lots of models to train and test our framework. In practice, a defender can safely claim its ownership with only 10 models and 20 fingerprints.

\mypara{Evaluation Metrics.}
The similarity between two fingerprints are defined as the cosine similarity of their representation vectors projected by the trained encoder.
The similarity between two models is the average similarities on all generated fingerprints. 

\begin{table}[t!]
\centering
\caption{Mean and STD of model similarities between the victim model and suspect model and p-value (lower is better) using 20 fingerprints (FMNIST).}
\vspace{-3mm}
\resizebox{.98\hsize}{!}{%
\begin{tabular}{cc|lll|lll}
\hline
\multicolumn{1}{l}{}                        & \multicolumn{1}{l|}{} & \multicolumn{3}{c|}{Piracy Model}           & \multicolumn{3}{c}{Homologous Model}        \\ \hline
\multicolumn{2}{l|}{Architecture \& Optimizer}                      & Mean   & STD    & P value                   & Mean                     & STD    & P value \\ \hline
\multicolumn{1}{c|}{\multirow{3}{*}{Arc B}} & SGD                   & 0.9990 & 0.0012 & 0.0                       & 10\textasciicircum{}(-5) & 0.0009 & 1.0     \\
\multicolumn{1}{c|}{}                       & Adam                  & 0.9974 & 0.0048 & 0.0                       & 0.0858                   & 0.2613 & 0.9167  \\
\multicolumn{1}{c|}{}                       & Rmsprop               & 0.9964 & 0.0060 & 0.0                       & 0.0002                   & 0.0015 & 1.0     \\ \hline
\multicolumn{1}{c|}{\multirow{3}{*}{Arc C}} & SGD                   & 0.9322 & 0.1882 & 10\textasciicircum{}(-15) & 0.0009                   & 0.0149 & 1.0     \\
\multicolumn{1}{c|}{}                       & Adam                  & 0.9959 & 0.0050 & 0.0                       & 0.0185                   & 0.0842 & 1.0     \\
\multicolumn{1}{c|}{}                       & Rmsprop               & 0.9734 & 0.0927 & 0.0                       & 0.0074                   & 0.0417 & 1.0     \\ \hline
\multicolumn{1}{c|}{\multirow{3}{*}{Arc D}} & SGD                   & 0.9980 & 0.0027 & 0.0                       & 0.1082                   & 0.3010 & 0.8445  \\
\multicolumn{1}{c|}{}                       & Adam                  & 0.9972 & 0.0036 & 0.0                       & 0.1024                   & 0.2492 & 0.5663  \\
\multicolumn{1}{c|}{}                       & Rmsprop               & 0.9977 & 0.0030 & 0.0                       & 0.0295                   & 0.0938 & 0.6942  \\ \hline
\multicolumn{1}{c|}{\multirow{3}{*}{Arc E}} & SGD & 0.8821 & 0.203 & 10\textasciicircum{}(-16) & 0.0377 & 0.1360 & 0.9923 \\
\multicolumn{1}{c|}{}                       & Adam                  & 0.9515 & 0.1392 & 0.0                       & 0.0010                   & 0.0036 & 1.0     \\
\multicolumn{1}{c|}{}                       & Rmsprop               & 0.9491 & 0.1450 & 0.0                       & 0.0002                   & 0.0015 & 1.0     \\ \hline
\end{tabular}
}

\label{tab:evl:hypotest1} 
\end{table}

\subsection{Fingerprint Identification and Matching }

\autoref{fig:eval:cdf} reports the results of similarity for piracy and homologous fingerprints with different network architectures. 
The CDF of similarity distribution is calculated during each 0.1 interval. For $x$ in X-axis, the Y-axis is the percentage of models that have similarity smaller than $x$. 

We conclude that \textbf{1)} Similarities of piracy fingerprints and homologous fingerprints distribute differently. The former gathers near 1 whereas the latter gathers near zero. \textbf{2)} Our encoder has good generalizability in the sense that the similarity gaps exist regardless of models' architectures. For all types of architectures, a large amount (\eg, $100 \%$ for CIFAR10) of fingerprints have similarities above $0.8$ whereas for homologous models, only a small portion of fingerprints (\eg, $20 \%$ for CIFAR10 ) have similarity above $0.4$. 
However, the similarity gap does vary between architectures. The encoder performs the best on architecture that is trained on (\eg, Arc.A for FMNIST and ResNet34 for CIFAR10). \textbf{3)} The variance of homologous fingerprints are larger than that of piracy models (\ie, in CIFAR10, the largest similarity gap between architectures is 0.9 versus 0.0 with CDF granularity equals 0.1). This indicates that piracy model is restricted to a small subspace whereas homologous models lie in a larger subspace. 

\autoref{tab:evl:hypotest1} and \autoref{tab:evl:hypotest2} show the average similarities between suspect models and the victim. The largest similarity gaps between piracy and homologous models for three datasets are $0.99$, $0.95$, $0.58$, respectively. The smallest gaps are still over $0.77$, $0.69$, $0.43$. TinyImageNet has the smallest similarity gap because its lower SOTA accuracy results in worse extraction performance. 
We will study the influence of extraction performance on our framework in ablation study.

\mypara{Hypothesis Tests.}
In practice, defenders can 
adopt Two-Sample $t$-Test to safely verify the suspect model in less than 20 fingerprints.
Formally, let $\Omega$ and $\Omega_{homo}$ be two sets of fingerprint similarities calculated from suspect models and homologous models. We define the null hypothesis as:
$    \mathcal{H}_0 : \mu  < \mu_{homo}, \;  \text{where} \; \mu = \overline{\Omega},  \;\; \mu_{homo} = \overline{\Omega}_{homo}. $
The $t$-Test will either reject $H_0$ with a controllable significance level $\alpha$ to claim a piracy model, or give inconclusive results.  

The  P value column in \autoref{tab:evl:hypotest1} and \autoref{tab:evl:hypotest2} is calculated using $t$-Test performed on $20$ randomly sampled fingerprints. We repeat this $t$-Test for 30 times and present the average value. By setting the predefined significance level $\alpha$ to 0.05, we successfully reject $H_0$ for all the piracy models, giving a detect success rate equals $100 \%$

\mypara{Comparison to Existing Methods.}
We use the area under the ROC curve (AUC) to measure the performance of our work. Compared with prior work (IPGuard~\cite{DBLP:conf/asiaccs/CaoJG21}), the performance of that is the AUC of 0.83, 0.75, 0.61 in three datasets (FMNIST, CIFAR10 and T-ImageNet), 
our framework exceeds theirs and achieves the AUC of 1.0, 1.0, 0.98.

\begin{table}[t!]
\centering
\caption{\small Mean and STD of model similarities between the victim and suspect model and p-value (lower is better) using 20 fingerprints (CIFAR10 \& TinyImageNet).}
\vspace{-3mm}
\resizebox{.98\hsize}{!}{%
\begin{tabular}{c|llll|llll}
\hline
\multicolumn{1}{l|}{}      & \multicolumn{4}{c|}{CIFAR10}                        & \multicolumn{4}{c}{TinyImageNet} \\ \hline
Archi & \multicolumn{1}{c}{Type} & \multicolumn{1}{c}{Mean} & \multicolumn{1}{c}{STD} & \multicolumn{1}{c|}{P value} & Type & Mean & STD & P value \\ \hline
\multirow{2}{*}{ResNet}    & \multicolumn{1}{c}{Piracy} & 0.9945 & 0.0032 & 0.0  &  Piracy & 0.7463      &   0.0955     &  10\textasciicircum{}(-14)       \\
                           & Homo                       & 0.0476 & 0.0156 & 1.0  &    Homo & 0.2500    & 0.1462       &    0.5278       \\ \hline
\multirow{2}{*}{VGG}       & Piracy                     & 0.9917 & 0.0050 & 0.0  &  Piracy & 0.7568      &   0.0937     &    10\textasciicircum{}(-16)        \\
                           & Homo                       & 0.1950 & 0.0920 & 0.99 &     Homo & 0.2602   &  0.1530      &    0.1574    \\ \hline
\multirow{2}{*}{GoogLeNet} & Piracy                     & 0.99   & 0.0066 & 0.0  &  Piracy      & 0.7280       &    0.1621    & 10\textasciicircum{}(-9)      \\
                           & Homo                       & 0.2984 & 0.1682 & 0.69 &    Homo    & 0.2404       &  0.1440      &  0.8277     \\ \hline
\multirow{2}{*}{DenseNet}  & Piracy                     & 0.9924 & 0.0044 & 0.0  &   Piracy     & 0.8215       &    0.1002    &     10\textasciicircum{}(-15)  \\
                           & Homo                       & 0.0552 & 0.1955 & 1.0  &  Homo      & 0.2943       &  0.1656      & 0.7088      \\ \hline

\end{tabular}
}

\label{tab:evl:hypotest2} 
\end{table}

\subsection{Ablation Study} 
\label{sec:abltsty}
\begin{figure}[t]
    \centering
	\begin{subfigure}[b]{0.23\textwidth}
		\centering
		\includegraphics[width=\linewidth]{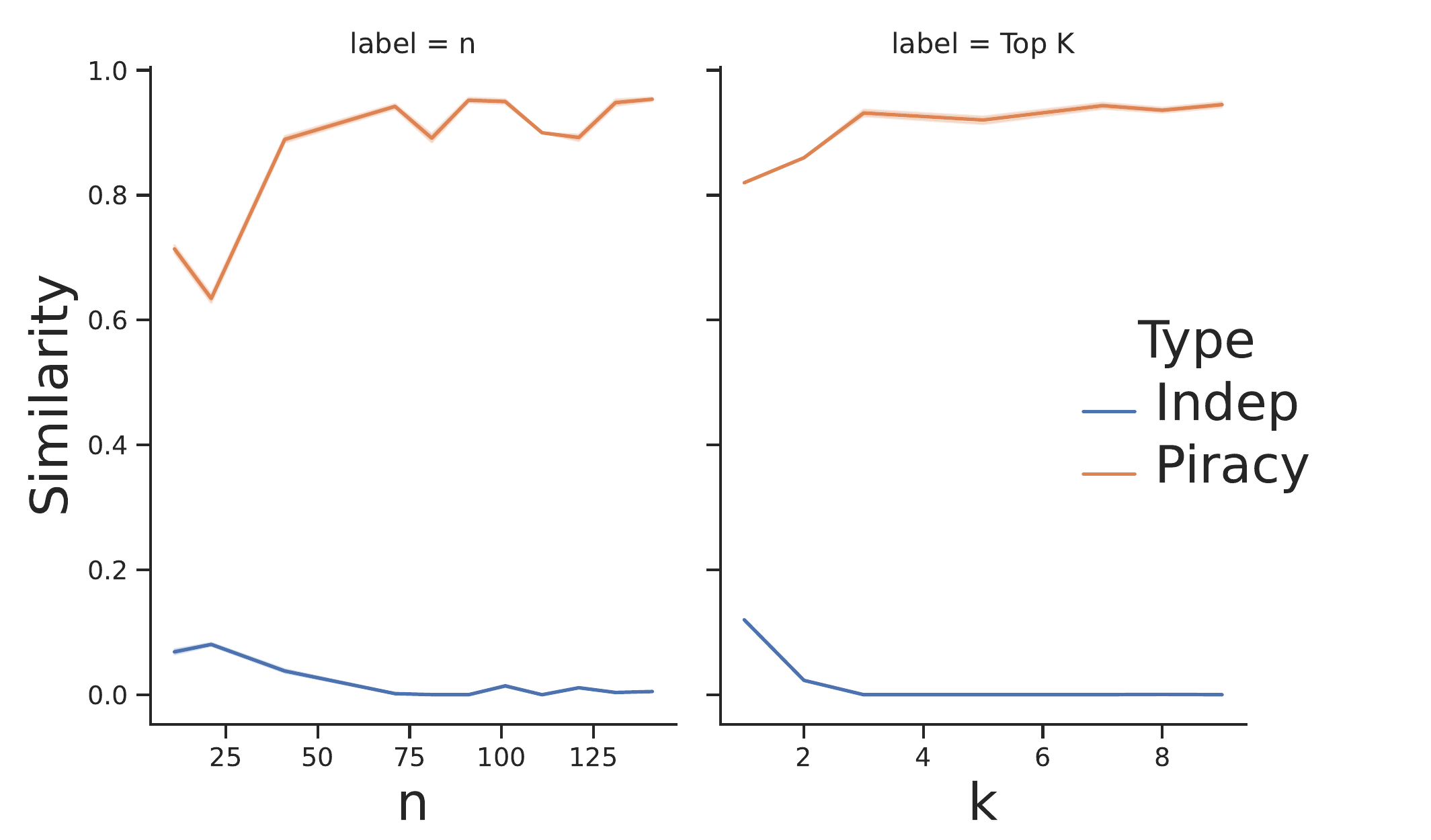}
		\label{fig:eval:abla_nk}
	\end{subfigure}
	\hfill
	\begin{subfigure}[b]{0.23 \textwidth}
    	\includegraphics[width=\linewidth]{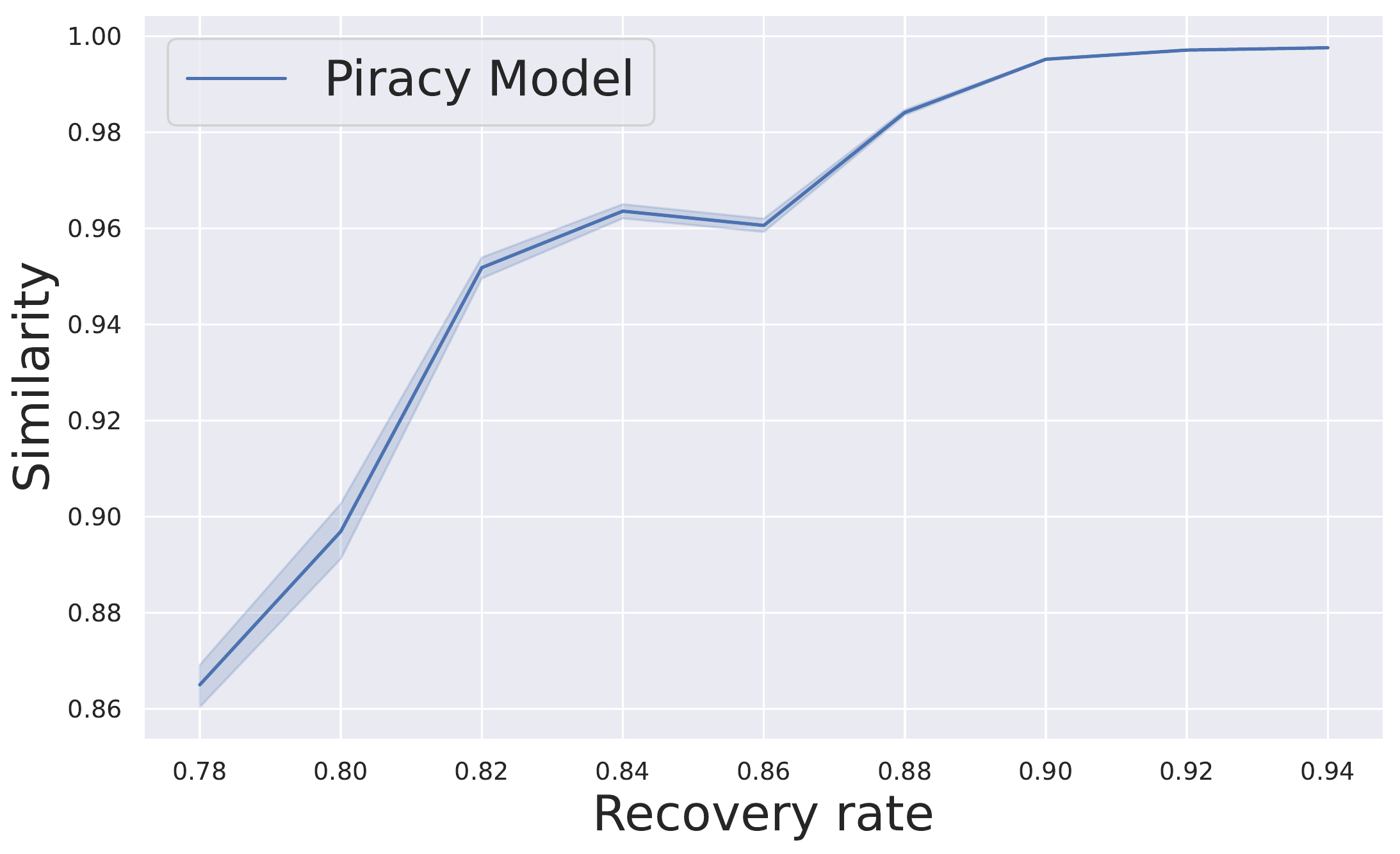}
    	\label{fig:eval:abla_cov}
    \end{subfigure}
    \vspace{-5mm}
    \caption{Ablation Study on parameters $n$, $k$ (left) and Recovery rate (right).}
	\label{fig:eval:abla}
	\vspace{-2mm}
\end{figure}

\mypara{Number of Datapoints $n$.}
Recall that our fingerprint generation function $\mathcal{F}(f, \mathbf{v}, \mathcal{X}_n)$ depends on a set of datapoints $\mathcal{X}_n$ where $n$ is the number of datapoints. 
As demonstrated in \autoref{fig:eval:abla}, with the increment of $n$, the similarity gap between piracy models and homologous models increases, until $n$ reaches a plateau at 70 datapoints. 
Larger $n$ means that fingerprint captures more decision boundaries and is more informative. 
In our experiments, we fix $n$ as 100 to balance the trade-off between effectiveness and efficiency.

\mypara{Top-K Confidence Values.}
We evaluate the performance of our framework when the suspect model only returns its top-k confidence values.
As demonstrated in Fig.~\ref{fig:eval:abla}, the similarity gap remains when $k = 1$, indicating that even hard-label can disclose global information of decision boundaries. The similarity gap is positively consistent with $k$ and becomes stable at $3$.
We find the average sum of returned top-3 confidence scores equals 0.9996, meaning that there is not much information loss.

\mypara{Performance of Model Extraction.}
An attacker may stop early before reaching the optimal recovery rate during extraction. By varying the query and iteration numbers, we obtain 20 models for each interval of length $0.02$ of recovery rate in $[0.78, 0.94]$. \autoref{fig:eval:abla} shows that when the recovery rate is $0.78$, our framework can still detect the piracy with a high similarity $0.88$. 
One explanation is when performing model extraction, the piracy models roughly form a decision boundary globally aligned with the victim, which enables our detection. Then it refines its local gradients which improves recovery rate and increases similarity.

\begin{figure}
    \centering
    
    \begin{subfigure}[b]{0.23\textwidth}
		\includegraphics[width=\linewidth]{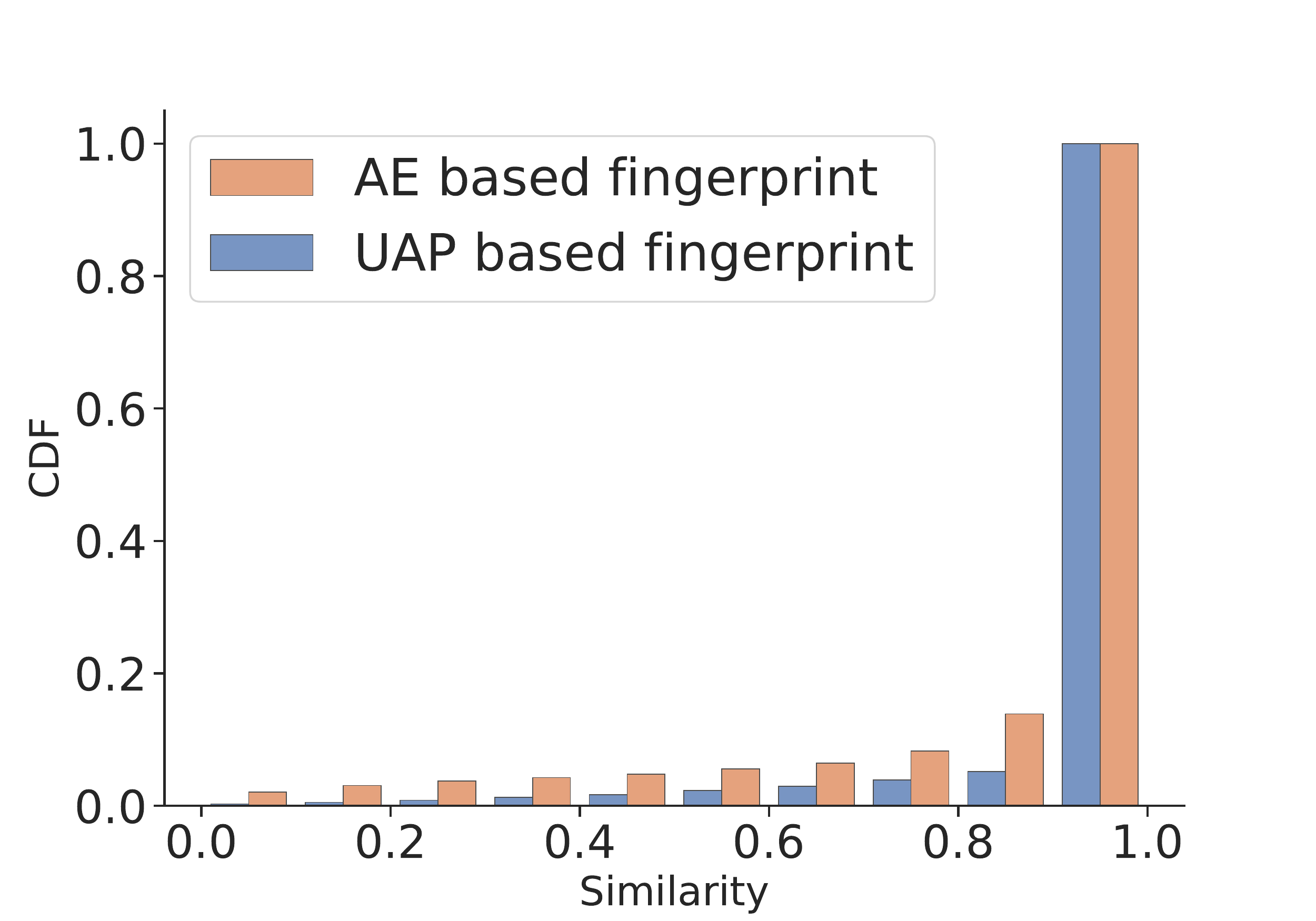}
		\caption{Sim CDF between $\vmodel{u}$ and $\pmodel{u}$ (FMNIST)}
		\label{fig:eval:cdf_sim_ind_ablation}
	\end{subfigure}
	\hfill
	\begin{subfigure}[b]{0.23\textwidth}
		\includegraphics[width=\linewidth]{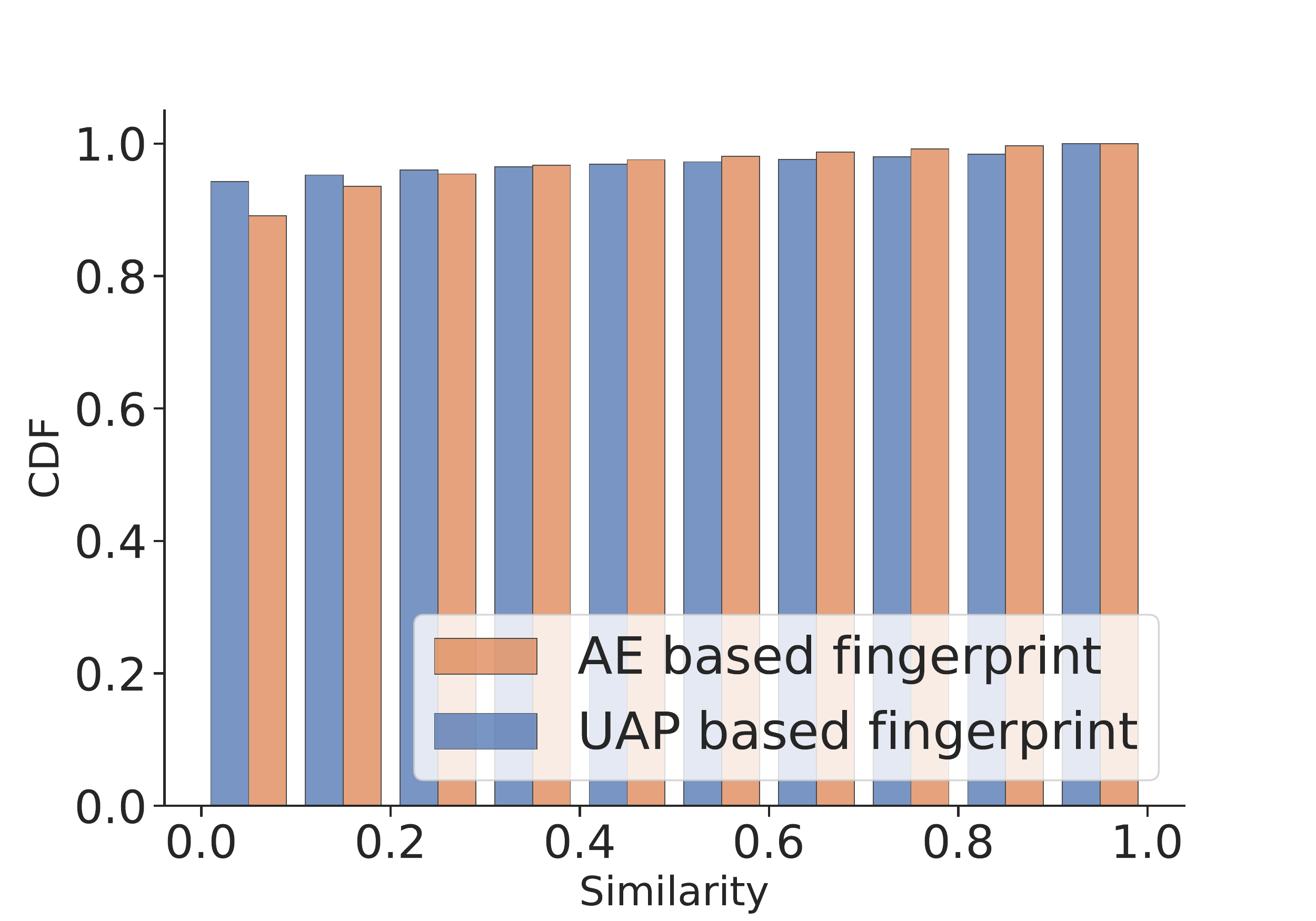}
		\caption{Sim CDF between $\vmodel{u}$ and $\vmodel{v}$ (FMNIST)}
		\label{fig:eval:cdf_sim_ind_ablation}
	\end{subfigure}
    \caption{The similarity score given by the contrastive encoder in UAP-based  and adversarial example based fingerprints. }
    \label{fig:abls_uap_ae_fmst}
    \vspace{-4mm}
\end{figure}

\mypara{Universal \textit{vs.} Local Adversarial Perturbations.}
To compare the information capture capabilities of UAP and AP, We replace UAP by AP and re-conduct our experiments. Specifically, the fingerprint is now 
$\mathcal{F}_{ap}(f, (\mathbf{x}_1,\cdots,\mathbf{x}_n) ) = [f(\mathbf{x}_1), f(\mathbf{x'}_1),\cdots,f(\mathbf{x}_n), f(\mathbf{x'}_n)$],
where $\mathbf{x'}$ is the AP of $\mathbf{x}$ crafted by DeepFool~\cite{DBLP:conf/cvpr/Moosavi-Dezfooli16} ($\epsilon = 22$). The perturbation norm of $\mathbf{x'}$ approximates UAP, other settings is unchanged. 
\autoref{fig:abls_uap_ae_fmst} shows the similarity score given by the contrastive encoder. We observe that for AP, the similarities of piracy models are less concentrated in 1 and that of homologous models are less concentrared in 0. This indicates AP based fingerprints have worse performance than UAP based one.

We also study the influence of the usage of contrastive loss and overlapping rate between homologous' and victim's datasets. See {\color{magenta}supplementary material} for details.

\subsection{Resistance against Model Modifications}
A smart attacker may deliberately modify the piracy copy in order to evade the detection,  
we evaluate the robustness of our framework against four post-processing techniques on FMNIST dataset . 

\mypara{Fine-tuning.}
Fine-tuning involves continuing training piracy models using additional data. In our experiment, 
We fine-tune piracy models on datapoints sampled from test dataset for $10$ iterations.

\begin{figure}[t]
    \centering
	\begin{subfigure}[b]{0.233\textwidth}
		\includegraphics[width=\linewidth]{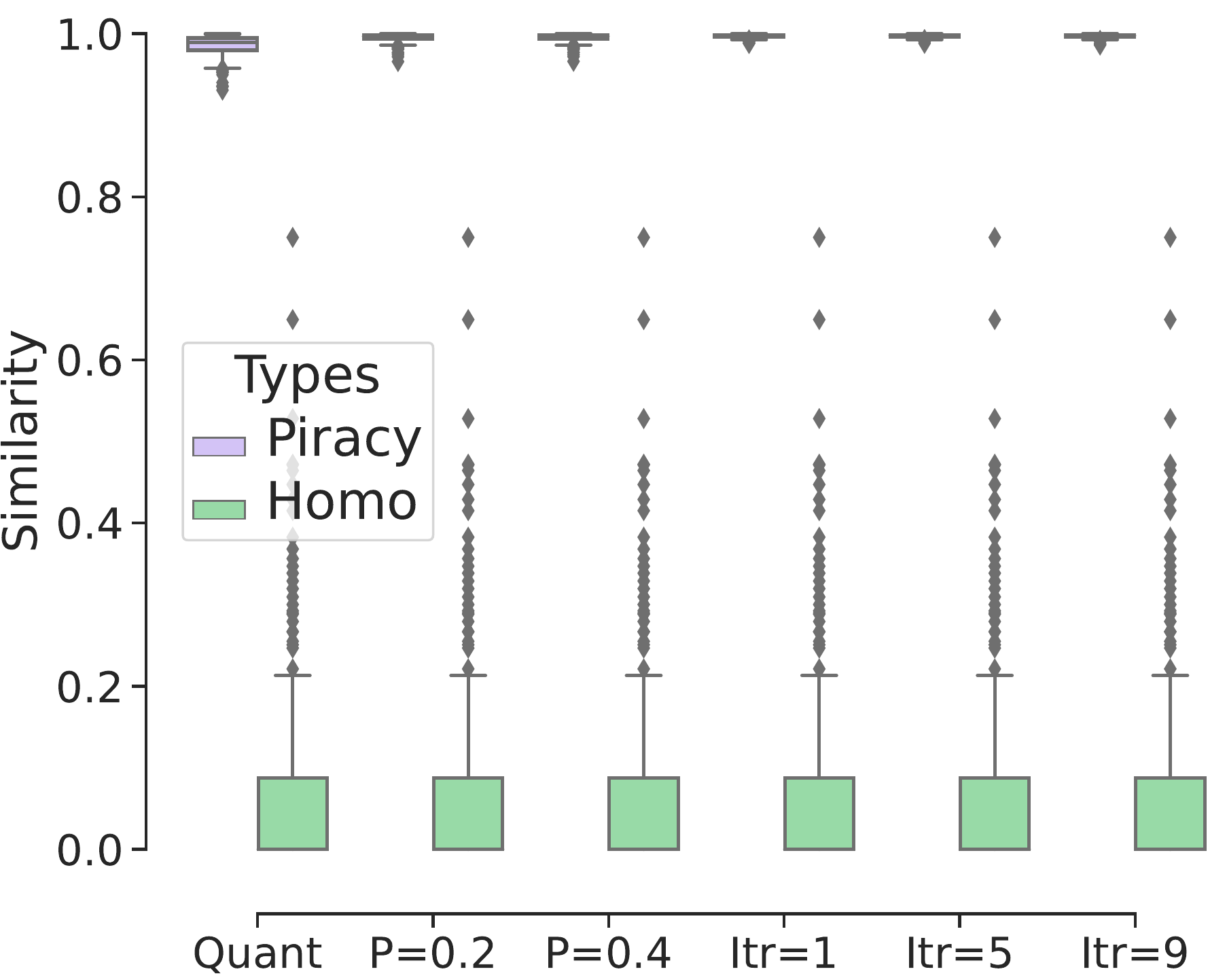}
		\label{fig:eval:modify_ft}
	\end{subfigure}
	\hfill
	\begin{subfigure}[b]{0.22\textwidth}
		\includegraphics[width=\linewidth]{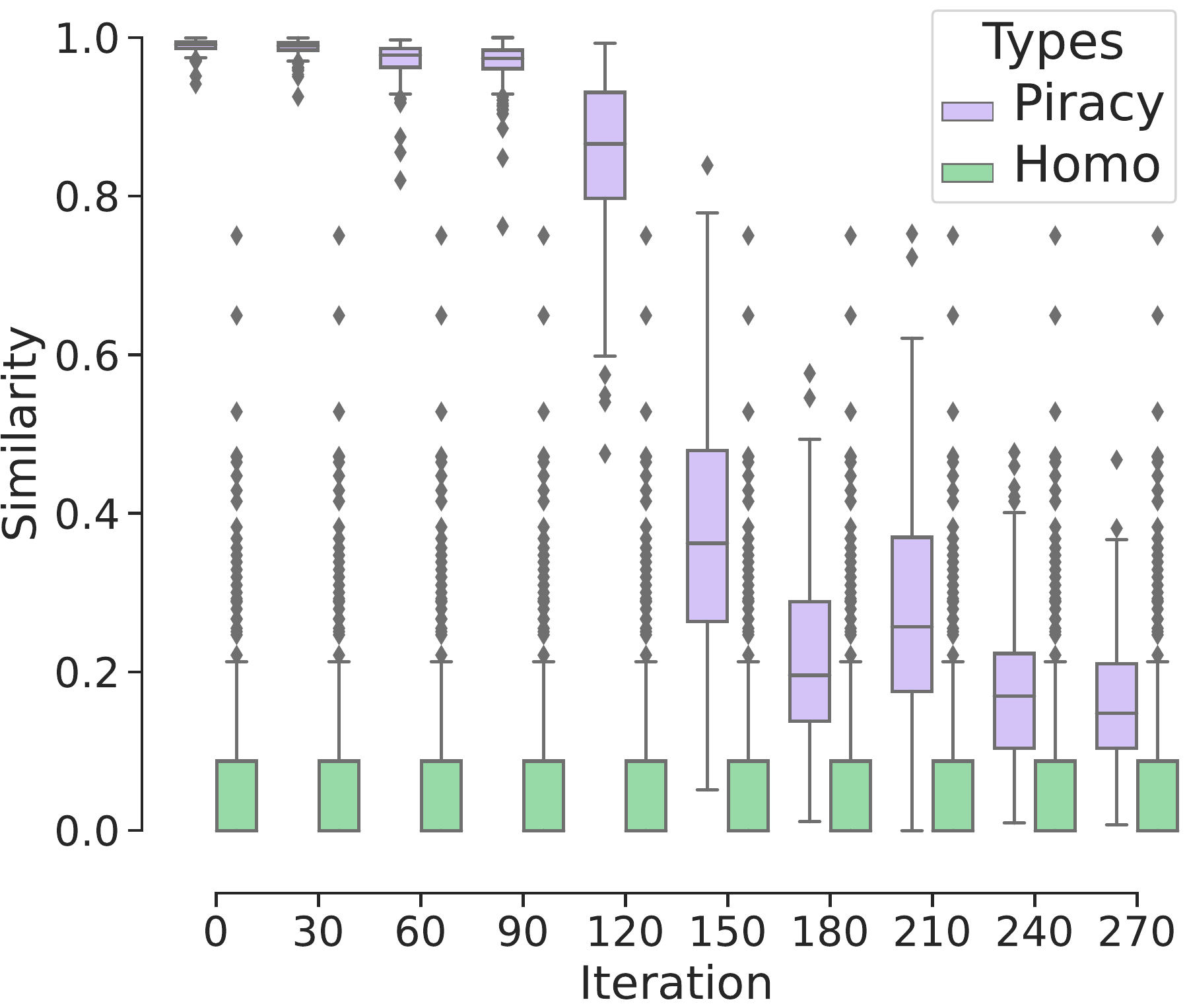}
		\label{fig:eval:modify_advtrain}
	\end{subfigure}
	\vspace{-5mm}
	\caption{Resistance against model modifications. Similarity distribution of piracy models after quantization, prune and fine tuning (left) and adversarial training (right). On the left, purple boxes assemble on top because their similarities achieve 1.0.}
	\label{fig:eval:resisi_md_mdfy}
	\vspace{-4mm}
\end{figure}

\mypara{Pruning \& Quantization.}
Pruning~\cite{DBLP:conf/iclr/ZhuG18} and quantization~\cite{DBLP:journals/corr/HanMD15} are two usual techniques to compress models and reduce memory while preserve model's functionality. In our experiment we choose pruning rate in $[0.2, 0.6]$ and we convert models from FP32 into INT8.
As \autoref{fig:eval:resisi_md_mdfy} (left) demonstrated, for the fine-tuning, quantization and pruning, the variation of similarities of piracy models are small. All modified piracy models by those three techniques still have similarities more than $0.88$. We hypothesize that they have little effect on the model's decision boundaries. 

\mypara{Adversarial Training.}
Adversarial training~\cite{DBLP:conf/iclr/MadryMSTV18} aims to promote the robustness of models intrinsically. We train the piracy models for a maximum 270 adversarial iterations. In each iteration, we craft 128 adversarial examples using DeepFool~\cite{DBLP:conf/cvpr/Moosavi-Dezfooli16} as new datapoints. 

As shown in \autoref{fig:eval:resisi_md_mdfy} (right), when the iteration is larger than 120, the similarity continues to drop from $0.99$ to $0.89$ but still remains high. This is because the adversarial training will continuously shape the model's decision boundaries by pushing the decision boundary towards adversarial examples.
The similarity gap is imperceptible after 270 adversarial training iterations with the model utility drops $0.17$. So the attacker faces a dilemma to sacrifice the stolen model's utility for evading our detection framework.

\section{Discussion}
In this paper, we propose a novel framework against model extraction attacks 
based on the subspace dependency of UAPs of the  victim model and piracy model. 
Incorporating with contrastive learning, we project victim model close to piracy models and far away from homologous models.
Evaluation on three benchmark datasets show that our framework is highly effective, general and robust. 

\mypara{Limitations.}
Our approach can be improved by making a better effectiveness-efficiency trade-off as the information in fingerprints relates to query number. We leave the overhead of training data preparation of encoder as another future work. Our discovery on transferability of encoder suggests a potential solution (see details in {\color{magenta}supplementary}).

{\small
\mypara{Acknowledgements.}
Z. Peng, S. Li, and H. Zhu were partially supported by the National Key Research and Development Program of China under Grant 2018YFE0126000 and the National Natural Science Foundation of China under Grant 6213000013. M. Xue was partially supported by the Australian Research Council (ARC) Discovery Project (DP210102670) and COVID-19 Recognition Fund of The University of Adelaide.
}

\small{
\bibliographystyle{ieee_fullname}
\bibliography{main}
}

\newpage
% \part*{Supplementary Materials}
\appendix
\begin{center}
\section*{Supplementary Material}
\end{center}

In this Supplementary Material, we provide details and results omitted in the main text.
\begin{itemize}[itemsep=0pt,topsep=2pt,leftmargin=12pt]
    \item \autoref{sec:lap}: local variations of decision boundaries during model extraction (\S~4.2 of the main paper)
    \item \autoref{sec:augmentation}: effectiveness of multi-views fingerprints augmentations (\S~4.3 of the main paper)
    \item \autoref{sec:ablation}: additional ablation studies (\S~5.3 of the main paper).
    \item \autoref{sec:all_model_acc}: additional results of model accuracy. (\S~5.2 of the main paper)
    \item \autoref{sec:transfer}: additional analyses on the transferability of encoder. (\S~6 of the main paper)
     \item  \autoref{sec:ethical}: discussion on ethical issues. (\S~6 of the main paper)
\end{itemize}

\section{Local Variations of Decision Boundaries During Model Extraction}
\label{sec:lap}
In \S~4.2 of the main paper, the visualization of fingerprints generated using UAP and Local Adversarial Perturbation (LAP) conducted on FashionMNIST dataset shows that LAP fingerprints are much less distinguishable than UAP fingerprints. This is because UAP based fingerprints capture the global information of decision boundaries, which is robust to model extraction process, as demonstrated in \S~4.1. In this section, we will show that the local information representing by LAP based fingerprints is much less robust because decision boundaries vary significantly during model extraction. 

\begin{figure}[t]
    \centering
    \includegraphics[width=0.85\linewidth]{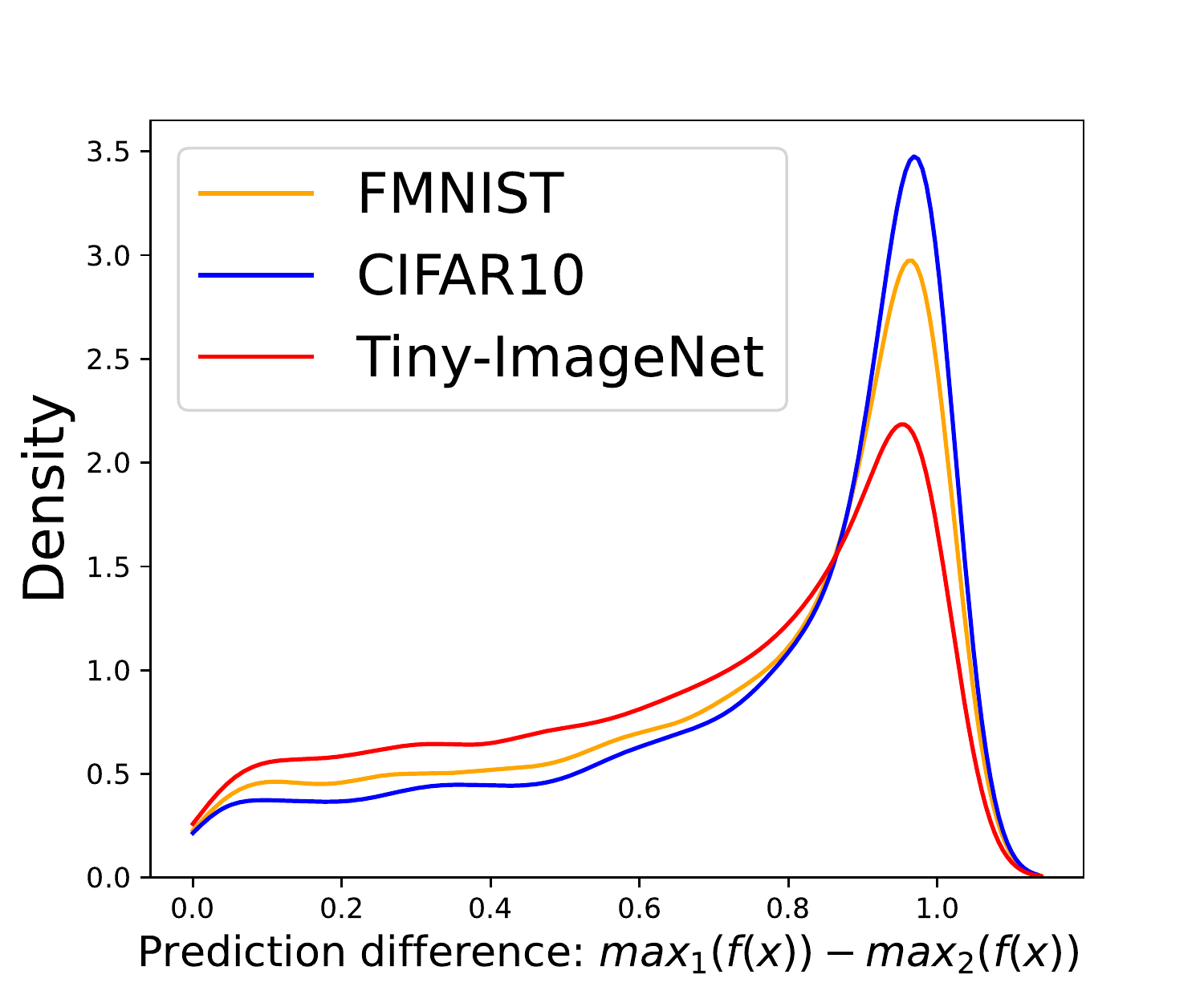}
    \caption{Distributions of prediction gap of borderpoints on piracy models.}
    \label{fig:supple:bordergap}
    \vspace{-2mm}
\end{figure}

To show the variation, we leverage a special kind of datapoints named borderpoints which are datapoints that lie on the decision boundary (\ie $\{ \mathbf{x} \, | \, argmax_1(f(\mathbf{x})) - argmax_2(f(\mathbf{x})) < 1e^{-6} \}$. As the last layer of DNN is a compact space, we can easily find such border points using dichotomy (\ie for any $\mathbf{x_i}$ in  class $C_i$ and any $\mathbf{x_j}$ in  class $C_j$,  there exists $ \lambda \in [0,1]$ that $ \lambda * \mathbf{x_i} + (1 - \lambda) * \mathbf{x_j}$ is a borderpoint). These borderpoints are used to query the piracy model of $f$ and \autoref{fig:supple:bordergap} reports the differences between the largest confidence value and the second largest confidence value of piracy models on borderpoints. 
As shown in \autoref{fig:supple:bordergap}, the gap between the confidence value of the most probable class and the value of the second probable class is significant, indicating that borderpoints are far away from decision boundaries of piracy models. In this way, we show the variance of decision boundaries during model extraction quantitatively.

\section{Effectiveness of Multi-views Fingerprints Augmentations}
\label{sec:augmentation}
In \S~4.3 of the main paper, we propose a data augmentation strategy which forms multiple views for one fingerprint and then use them to train the contrastive encoder. In this section, we verify the effectiveness of this strategy by proving that among all the positives (\ie, samples belongs to the same class), those generated views of a specific fingerprint are the most similar to itself.  

\begin{figure}[t] 
    \centering
    \begin{subfigure}[t]{0.23\textwidth}
        \centering
        \includegraphics[width=0.99\linewidth]{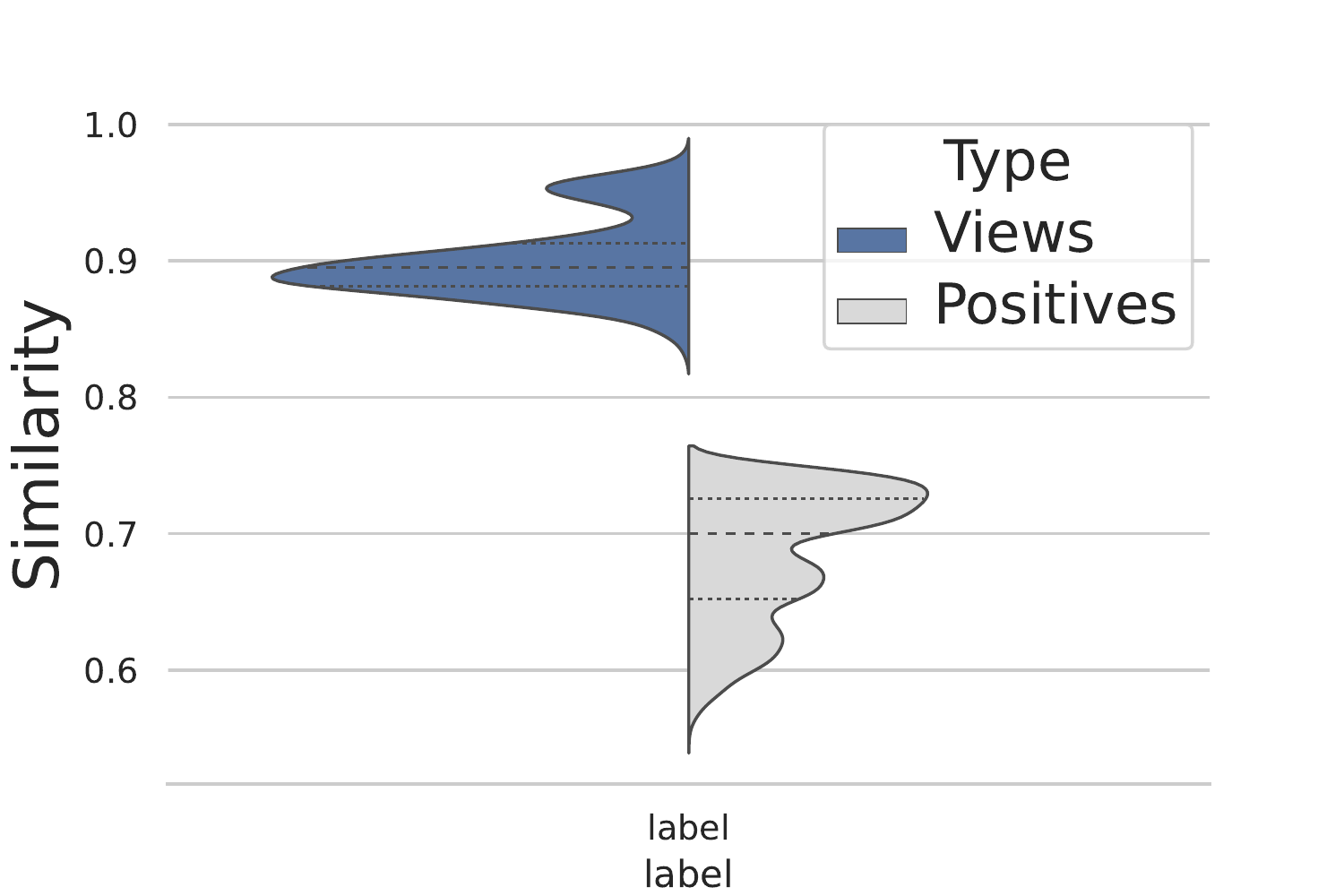}
        \vspace{-0.2cm}
        \caption{Similarity distribution}
        \label{fig:viesim_1}
    \end{subfigure}
    \begin{subfigure}[t]{0.23\textwidth}
        \centering
        \includegraphics[width=0.99\linewidth]{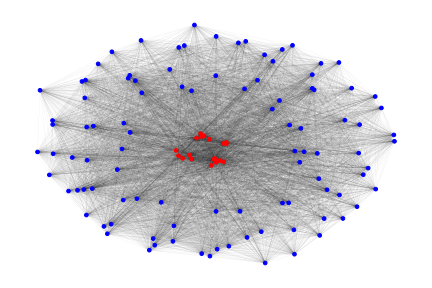}
        \vspace{-0.2cm}
        \caption{Visualization of fingerprints}
         \label{fig:viewsim_2}
    \end{subfigure}
    \caption{\small (a) The left curve is the similarity between a fingerprint and its views, the right curve is the similarity between a fingerprints and its positives excluding views; (b)Visualization of fingerprints: red points are views generated from a single fingerprint and blue points are other positives with red points.}
    %\vspace{-0.1cm}
    %\vspace{-0.cm}
\end{figure}

The similarities $s$ between views and positives are measured by cosine similarities. \autoref{fig:viesim_1} reports the similarities between a pair of views and a pair of positives. We conclude that compared with positives, views are more similar with each other which is satisfied with our design choice. The graph in \autoref{fig:viewsim_2} shows the distances between a pair of fingerprints measured by the reciprocal of their similarities $d = (1 / s)$ which reveals the same results.

\section{Additional Ablation Studies}
\label{sec:ablation}
\subsection{Contrastive Loss}
\label{sec:abla_contra}
In this section, we aim at presenting the role of contrastive learning in encoder training. Recall that contrastive learning can distinguish the differences between homologous models and piracy models (\ie, the representation vector of homologous models will be distant from the victim model whereas that of piracy models will be close to the victim model). This distance property in the latent space allows us to stably verify the similarity between two models and detect piracy models with high confidence. 

\begin{figure}[t] 
    \centering
    \begin{subfigure}[t]{0.23\textwidth}
        \centering
        \includegraphics[width=0.99\linewidth]{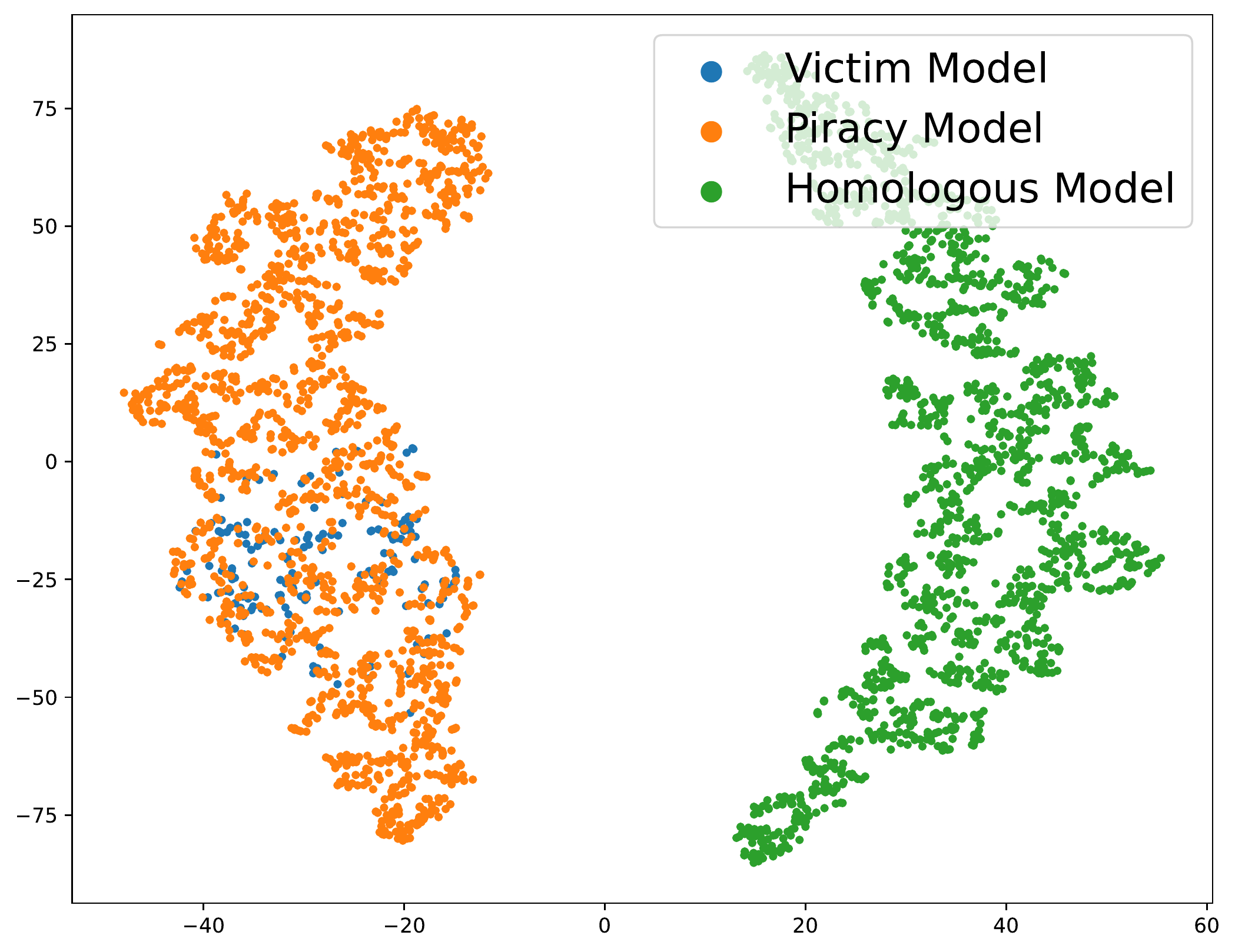}
        \vspace{-0.2cm}
        \caption{Contrastive encoder}
        \label{fig:visual_feat_L2_clean}
    \end{subfigure}
    \begin{subfigure}[t]{0.23\textwidth}
        \centering
        \includegraphics[width=0.99\linewidth]{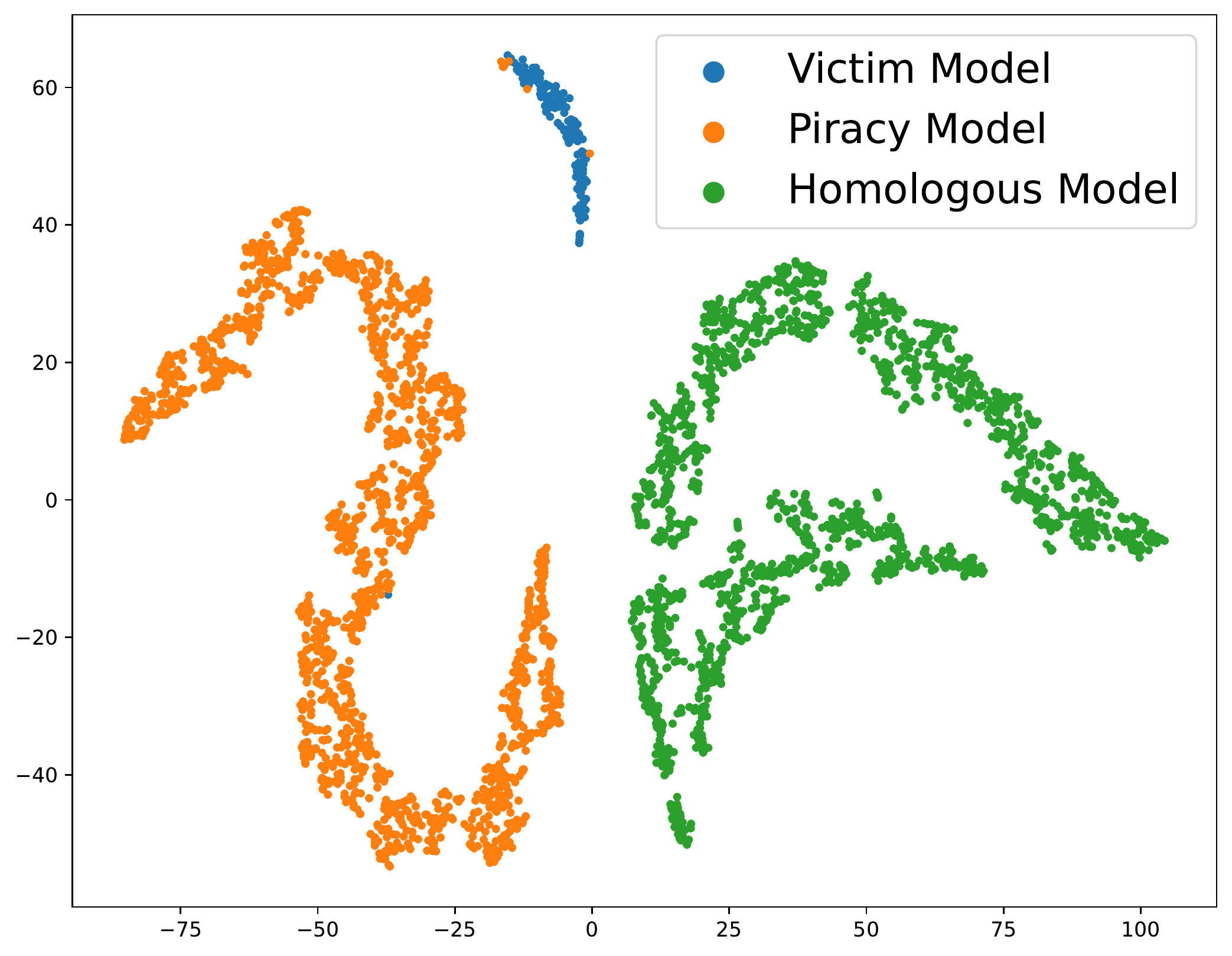}
        \vspace{-0.2cm}
        \caption{Non-contrastive encoder}
        \label{fig:visual_feat_L2_poison}
    \end{subfigure}
 
    \caption{t-SNE visualization of representation vector outputted by the encoder of three types of fingerprints (FMNIST). Contrastive learning based encoder (left) can better differentiate homologous models from piracy models and can better mix piracy models with the victim model than non-contrastive encoder (right).}
    \label{fig:visual_feature_space}
\end{figure}

To understand how well can contrastive learning based encoder differentiates models, we visualize the distribution of the representation vector of each fingerprint in the last layer of encoder and we compare the case of contrastive encoder with that of normal auto-encoder. \autoref{fig:visual_feature_space} is t-SNE visualization about the representation vector of fingerprints where each group of points with different colors represents one type of fingerprint. 
\autoref{fig:visual_feat_L2_clean} demonstrates the results of a contrastive learning based encoder. The representation vectors of the victim model (blue) are entangled with that of piracy models (orange) and lie on the left side of this hyper-sphere while the homologous models (green) lie on the opposite side. This is exactly what we expected because there exists an obvious decision boundary in \autoref{fig:visual_feat_L2_clean} that can easily split those three clusters into two parts.  
% The distances between clusters maintain on each pair of points. 
In contrast, \autoref{fig:visual_feat_L2_poison} shows the results of a non-contrastive encoder (\ie, auto-encoder). As we can see, although each type of fingerprint is separated (recall that UAP based fingerprint itself is separable without any post-processing), the distances of any two points from different clusters variant. Besides, the representation vectors of victim's fingerprints and that of piracy models do not have any overlap. We thus conclude that the contribution of contrastive learning is vital in achieving nearly perfect detection rate.    

\subsection{Datasets Overlapping}

In \S~5.2 of the main paper, we give the implementation details about the training process of victim model, homologous models and piracy models. In this section, we will report the influence of the overlapping rate between the dataset of victim models and the dataset of homologous models with respect to the performance of our verification mechanism. Note that we are only interested in the overlapping rate between $D_v$ and $D_{homo}$ rather than $D_v$ and the dataset of piracy models $D_{pir}$,  as the data augmentation technology used in the model extraction process makes it hard to measure their overlapping rate.

\begin{figure}[t] 
    \centering
    \includegraphics[width=0.85\linewidth, height=0.65\linewidth]{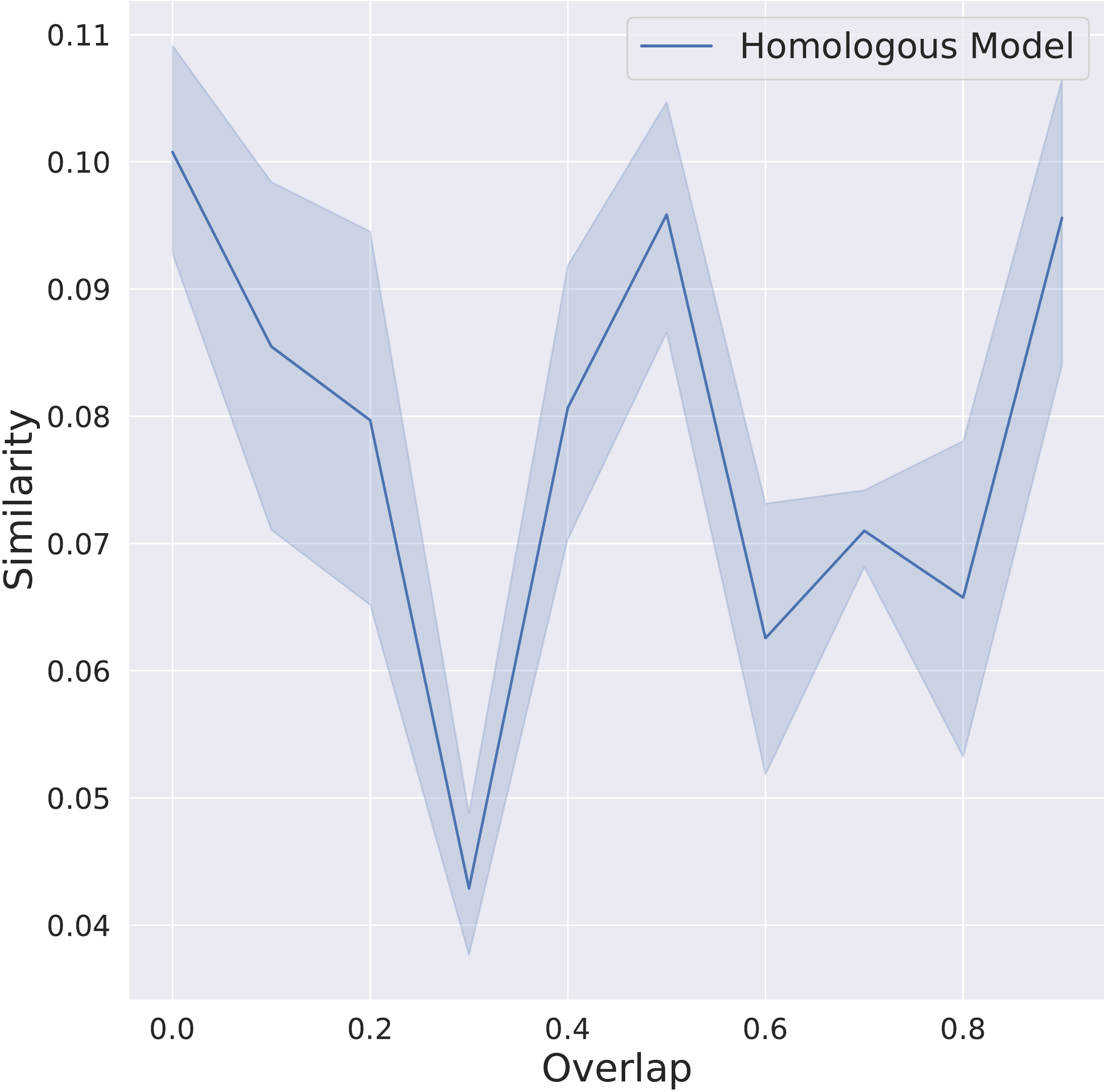}
    % \vspace{-0.2cm}
    \caption{Influence of the Overlapping rate between $D_v$ and $D_{homo}$ on similarities of homologous models and the victim model.}
     \label{fig:overlap}
\end{figure}

Intuitively, a homologous model trained on $D_{homo}$ which overlaps more with $D_v$ will assemble more the victim model and will be more difficult to be distinguished. A worth-trusty verification mechanism, however, need to be indifferent to such interference. To measure the effect of overlapping rate on our mechanism, we trained 20 homologous models with overlap rates ranging from $0$ to $0.9$ and calculated their similarities with $\vmodel{u}$. Our experimental results in \autoref{fig:overlap} on FMNIST show that the overlapping rate does not undermine the performance of our verification mechanism. Which indicates our approach is effectinve on different datasets overlaps.

\label{sec:abla_overlap}

\section{Model Accuracy} \label{sec:all_model_acc}
In this section, we present the accuracy on testset of all models used in our experiments in \autoref{tab:model_acc} as a supplement to \S~5.2 of the main paper.
% to the experiment setup.

\section{Transferability of Encoder}
\label{sec:transfer}
In \S~4.3 of the main paper, the defender needs to train several homologous models and several piracy models in order to train an encoder which satisfies the detection demand. To ease the burden of defenders, in this section, we aim to verify one hypothesis: can an encoder trained for a specific victim model $\vmodel{v_1}$ be used to protect another independent victim model $\vmodel{v_2}$.
Notice that $\vmodel{v_1}$ is independent from $\vmodel{v_2}$.

We evaluate this hypothesis on the FashionMNIST dataset. Specifically, given two independent victim models $\vmodel{v_1}$ and $\vmodel{v_2}$, \eg,  $\vmodel{v_2}$ is a homologous model of $\vmodel{v_1}$, we train an encoder $E_{v_1}$ for $\vmodel{v_1}$ to protect its ownership via our framework.
To test the transferability of this encoder, we additionally prepare 10 homologous models and 10 piracy models for $\vmodel{v_2}$ as well as its UAP. 
By far, we can generate three types of fingerprints for $\vmodel{v_2}$, \ie, the fingerprints of itself, its homologous models' fingerprints and its piracy models' fingerprints and we employ $E_{v_1}$ to project these fingerprints to the representation space.   

\begin{figure}[t] 
    \centering
    \begin{subfigure}[t]{0.23\textwidth}
        \centering
        \includegraphics[width=0.99\linewidth]{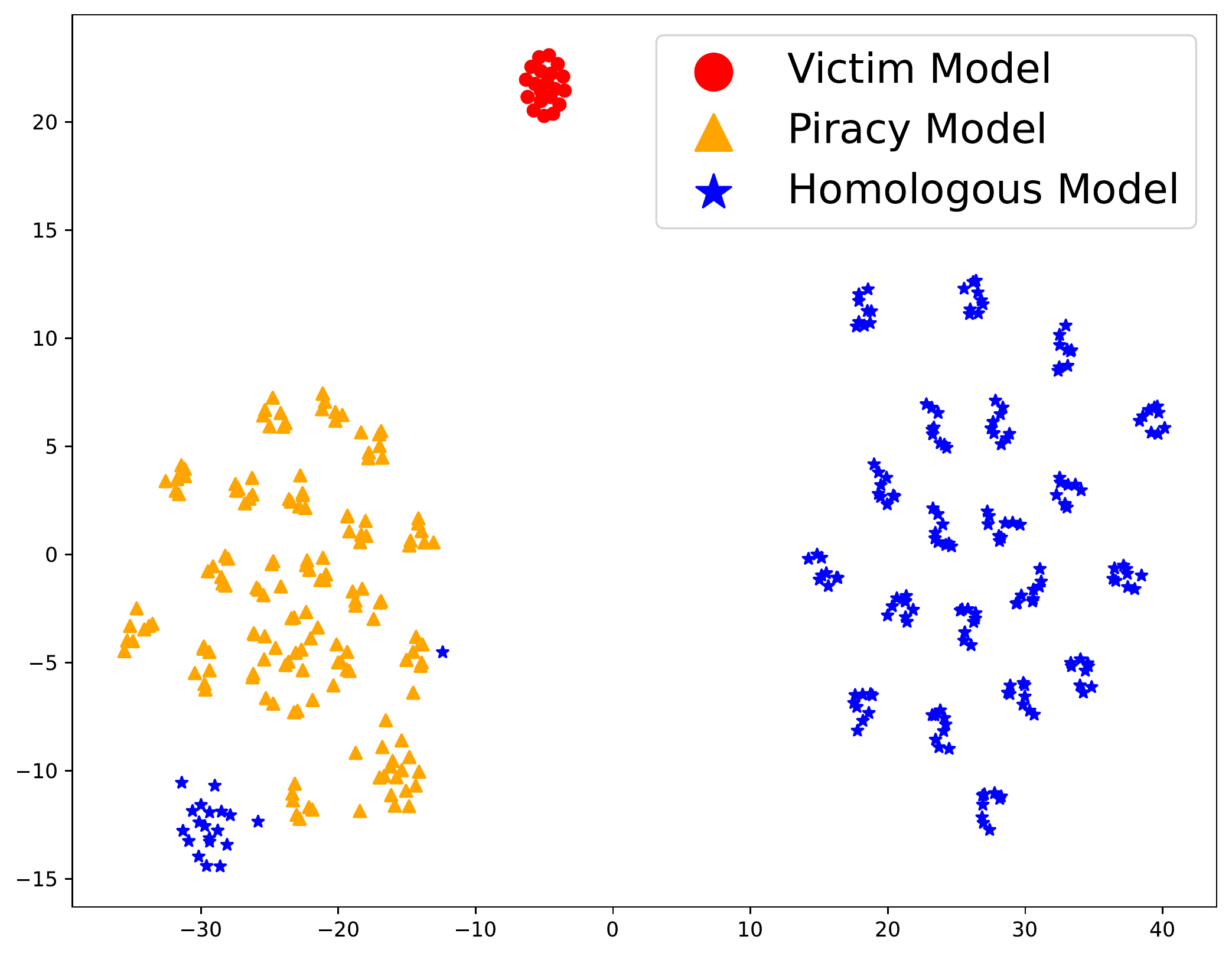}
        \vspace{-0.2cm}
        \caption{Fingerprints without encoder}
        \label{fig:visual_finger}
    \end{subfigure}
    \begin{subfigure}[t]{0.23\textwidth}
        \centering
        \includegraphics[width=0.99\linewidth]{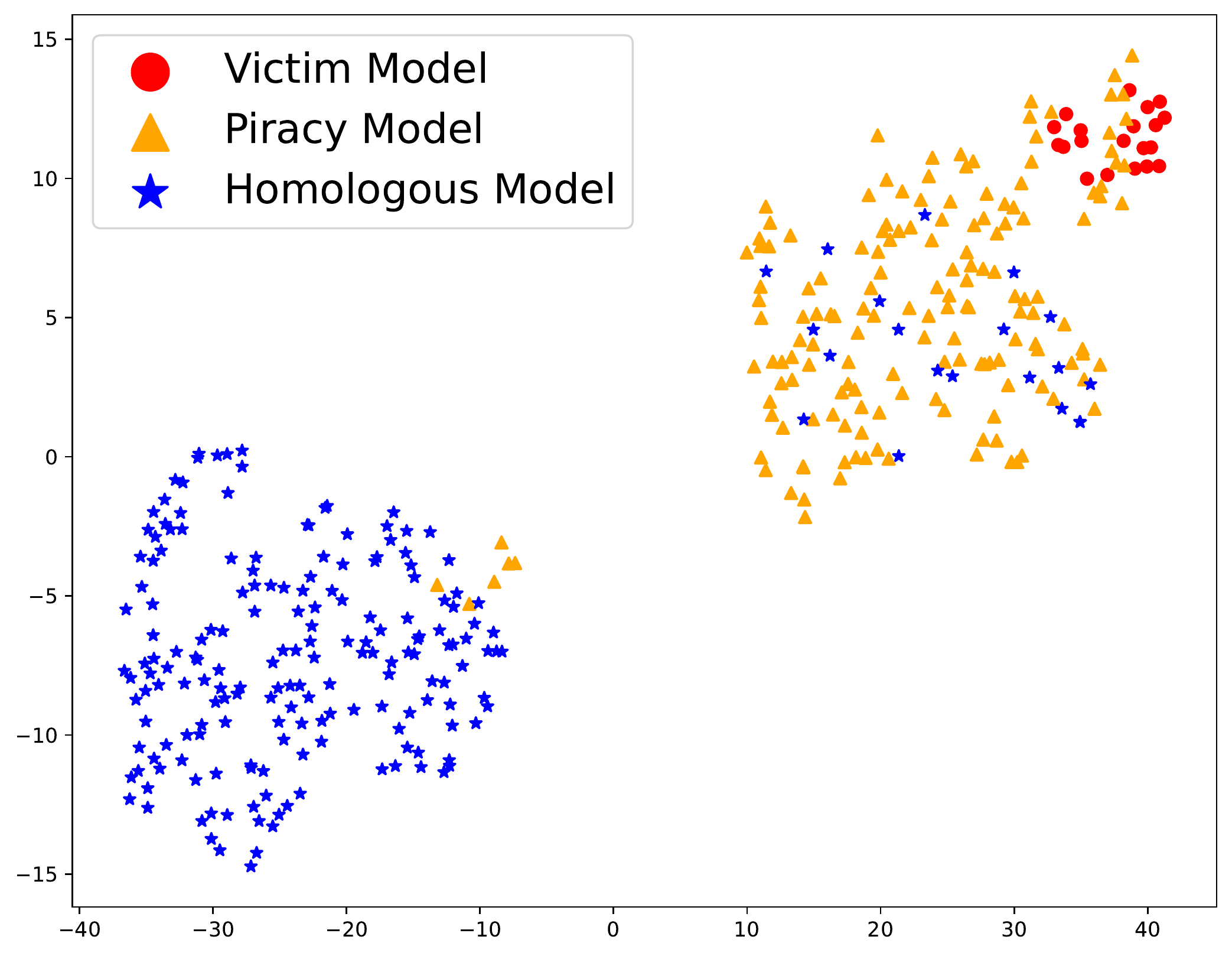}
        \vspace{-0.2cm}
        \caption{Representation vector outputed by encoder}
        \label{fig:visual_repre}
    \end{subfigure}
 
    %\vspace{-0.1cm}
    \caption{t-SNE visualization of fingerprints of three types of models associated with $\vmodel{v_2}$ and their representation vector outputed by the encoder which is associated with $\vmodel{v_1}$ (FMNIST).}
    \label{fig:visual_transfer}
    \vspace{-3mm}
\end{figure}

As demonstrated in \autoref{fig:visual_finger}, UAP based fingerprints are naturally separable, which unites with aforementioned experimental results. 
More importantly, as \autoref{fig:visual_repre} shown, after the projection of the encoder $E_{v_1}$, which is trained for another independent victim model, we observe that homologous fingerprints are indeed distant from the victim, whereas the piracy fingerprints are entangled with that of the victim. 
The distances of any two points belong to different clusters are maintained on most pairs of points. 
The average similarity $sim(f_{pir}, \vmodel{v_2})$ for measuring the IP violation equals $0.85$ and the average similarity $sim(f_{homo}, \vmodel{v_2})$ for independent models equals $0.33$. However, the similarity gap reduces slightly compared with the specific encoder that trained to protect itself (\ie, $\vmodel{v_2}$) and 
we claim that there is still space to improve the transferability of encoder in future work. 

\begin{table*}[t]
\caption{Accuracy of models of different architectures on different datasets. The values before brackets are the average and the values in brackets are STD. Arc.A and ResNet18 are chosen to be architecture of the victim model, so only one model of this architecture is generated.} 
\label{tab:model_acc}
\centering
\resizebox{.99\hsize}{!}{%
\begin{tabular}{c|ll|c|ll|ll}
\hline
\multicolumn{1}{l|}{} &
  \multicolumn{2}{c|}{FashionMNIST} &
  \multicolumn{1}{l|}{} &
  \multicolumn{2}{c|}{CIFAR10} &
  \multicolumn{2}{c}{TinyImageNet} \\ \hline
Architecture &
  Normal Training &
  \multicolumn{1}{c|}{Extraction} &
  \multicolumn{1}{l|}{Architecture} &
  Normal Training &
  \multicolumn{1}{c|}{Extraction} &
  Normal Training &
  \multicolumn{1}{c}{Extraction} \\ \hline
Arc A & 0.8978         &                & ResNet18  & 0.8929         &                & 0.4768         &                \\ \hline
Arc B & 0.8562(0.0314) & 0.8786(0.0125) & ResNet34  & 0.8956(0.0034) & 0.8918(0.0010) & 0.4754(0.0582) & 0.4354(0.0040) \\ \hline
Arc C & 0.9102(0.0529) & 0.8733(0.0071) & VGG16     & 0.9232(0.0021) & 0.8860(0.0009) & 0.4952(0.0520) & 0.4392(0.0076) \\ \hline
Arc D & 0.8648(0.0143) & 0.8698(0.0057) & GoogLeNet & 0.9185(0.0057) & 0.9075(0.0007) & 0.4561(0.0298) & 0.3520(0.0232) \\ \hline
Arc E & 0.8805(0.0223) & 0.8845(0.0067) & DenseNet  & 0.9185(0.0057) & 0.8962(0.0012) & 0.5406(0.0305) & 0.4114(0.0105) \\ \hline
\end{tabular}
}
\end{table*}

\section{Ethical Considerations}
\label{sec:ethical}
This work is mainly a defense paper against model extraction attacks and it is hardly misused by ordinary people. In the worst case, an honest-but-curious adversary may adopt this technique to involve a normal MLaaS provider in a lawsuit that is destined to lose. This concern can be eliminated by deploying a trustworthy third party to audit the argument. 

This work does not collect data from users or cause potential harm to vulnerable populations. It may arise concerns that the query data used by the defender reveal certain information about membership of their training dataset.  Fortunately, in our work, the query data do not need to belong to the user's original dataset, which means that the defender can collect auxiliary data that fall in the problem domain as their query data.

The other concern is that, although our verification framework can achieve a nearly perfect accuracy with an AUC score of 1.0, we still need to pay attention to the negative impact caused by its false positive cases. Further evidence collected by social engineering can be applied as auxiliary evidence during the confirmation process to avoid the hardly appeared false positive cases mentioned above.  

\end{document}